\documentclass[preprint]{ptephy_v1}

\preprintnumber{XXXX-XXXX} 

\usepackage{url}

\newcommand{\Nuc}[3]{$^{#2}_{#3}{\rm #1}$}

\usepackage{framed,color}
\usepackage{hyperref}

\usepackage{siunitx}
\begin{document}

\title{First observation of a nuclear $s$-state of $\Xi$~hypernucleus, $^{15}_{\Xi}{\rm C}$}


\author[1*]{M.~Yoshimoto}
\author[2]{J.~K.~Ahn}    
\author[3]{B.~Bassalleck}
\author[4]{H.~Ekawa}
\author[1]{Y.~Endo}
\author[5]{M.~Fujita}
\author[6]{Y.~Han}
\author[7]{T.~Hashimoto}
\author[7]{S.~H.~Hayakawa}
\author[8]{K.~Hicks}
\author[1]{K.~Hoshino}
\author[9]{S.~Hoshino}
\author[10]{S.~H.~Hwang}
\author[7]{Y.~Ichikawa}
\author[11,12]{M.~Ichikawa}
\author[7]{K.~Imai}
\author[5]{Y.~Ishikawa}
\author[5]{H.~Kanauchi}
\author[4,13]{A.~Kasagi}
\author[2]{S.~H.~Kim}
\author[13]{S.~Kinbara}
\author[13]{P.~M.~Lin}
\author[14]{T.~L.~Ma}
\author[5]{K.~Miwa}
\author[15]{A.~T.~Moe}
\author[1]{Y.~Nagase}
\author[1,13]{K.~Nakazawa}
\author[7,11]{T.~Nanamura}
\author[7,11]{M.~Naruki}
\author[13]{A.~N.~L.~Nyaw}
\author[16,17]{J.~Pochodzalla}
\author[7]{H.~Sako}
\author[7]{S.~Sato}
\author[18]{M.~M.~Soe}
\author[13]{M.~K.~Soe}
\author[19]{J.~Y.~Sohn}
\author[20]{H.~Takahashi}
\author[20]{T.~Takahashi}
\author[5,7]{H.~Tamura}
\author[7]{K.~Tanida}
\author[13]{A.~M.~M.~Theint}
\author[1]{K.~T.~Tint}
\author[5,20]{M.~Ukai}
\author[7]{T.~O.~Yamamoto}
\author[2]{S.~B.~Yang}
\author[19]{C.~S.~Yoon}
\author[4,5]{J.~Yoshida}
\author[14]{D.~H.~Zhang}
\author[14]{Z.~Zhang}
\affil[1]{Faculty of Education, Gifu University, Gifu 501-1193, Japan}
\affil[2]{Department of Physics, Korea University, Seoul 02841, Korea}
\affil[3]{Department of Physics and Astronomy, University of New Mexico, Albuquerque, New Mexico 87131, USA}
\affil[4]{High Energy Nuclear Physics Laboratory, RIKEN, Wako 351-0198, Japan}
\affil[5]{Department of Physics, Tohoku University, Sendai 980-8578, Japan}
\affil[6]{Hefei Institutes of Physical Science, Chinese Academy of Sciences, Hefei 230031, China}
\affil[7]{Advanced Science Research Center, Japan Atomic Energy Agency, Tokai 319-1195, Japan}
\affil[8]{Department of Physics \& Astronomy, Ohio University, Athens, Ohio 45701, USA}
\affil[9]{Department of Physics, Osaka University, Toyonaka 560-0043, Japan}
\affil[10]{Korea Research Institute of Standards and Science, Daejeon 34113, Korea}
\affil[11]{Department of Physics, Kyoto University, Kyoto 606-8502, Japan}
\affil[12]{Meson Science Laboratory, RIKEN, Wako 351-0198, Japan}
\affil[13]{Graduate School of Engineering, Gifu University, Gifu 501-1193, Japan}
\affil[14]{Institute of Modern Physics, Shanxi Normal University, Linfen 041004, China}
\affil[15]{Department of Physics, Lashio University, Buda Lane, Lashio 06301, Myanmar}
\affil[16]{Helmholtz Institute Mainz, 55099 Mainz, Germany}
\affil[17]{Institut f\"ur Kernphysik, Johannes Gutenberg-Universit\"at, 55099 Mainz, Germany}
\affil[18]{Department of Physics, University of Yangon, Yangon 11041, Myanmar}
\affil[19]{Research Institute of Natural Science, Gyeongsang National University, Jinju 52828, Korea}
\affil[20]{Institute of Particle and Nuclear Studies, High Energy Accelerator Research Organization (KEK), Tsukuba 305-0801, Japan\email{yoshimoto@onsanai.com}}




\begin{abstract}%
Bound-systems of $\Xi^-$--$^{14}_{}{\rm N}$ are studied via $\Xi^-$ capture at rest followed by emission of a twin single-$\Lambda$ hypernucleus in the emulsion detectors.
Two events forming extremely deep $\Xi^-$ bound states were obtained by analysis of a hybrid method in the E07 experiment at J-PARC and reanalysis of the E373 experiment at KEK-PS.
The decay mode of one event was assigned as $\Xi^-+^{14}_{}{\rm N}\to^{5}_{\Lambda}{\rm He}$+$^{5}_{\Lambda}{\rm He}$+$^{4}_{}{\rm He}$+n. 
Since there are no excited states for daughter particles, the binding energy of the $\Xi^-$ hyperon, $B_{\Xi^-}$, in $^{14}_{}{\rm N}$ nucleus was uniquely determined to be 6.27~$\pm$~0.27~MeV. 
Another $\Xi^-$--$^{14}_{}{\rm N}$ system via the decay $^{9}_{\Lambda}{\rm Be}$ + $^{5}_{\Lambda}{\rm He}$ + n brings a $B_{\Xi^-}$ value, 8.00~$\pm$~0.77~MeV or 4.96~$\pm$~0.77~MeV, where the two possible values of $B_{\Xi^-}$ correspond to the ground and the excited states of the daughter $^{9}_{\Lambda}{\rm Be}$ nucleus, respectively. 
Because the $B_{\Xi^-}$ values are larger than those of the previously reported events (KISO and IBUKI), which are both interpreted as the nuclear $1p$ state of the $\Xi^-$--$^{14}_{}{\rm N}$ system,
these new events give the first indication of the nuclear $1s$ state of the $\Xi$ hypernucleus, $^{15}_{\Xi}{\rm C}$.
\end{abstract}

\subjectindex{D00, D14, D25, D29}

\maketitle

\section{Introduction}

Elucidation of the various baryon--baryon interactions is a goal of modern nuclear physics.
To construct a framework which generalizes the nuclear force, studies are being conducted on interactions between the octet baryons of flavor $SU(3)$ which contain strange quarks ($s$) in addition to up ($u$) and down ($d$) quarks. 
Although the mass of the strange quark is smaller than the QCD scale parameter ($\Lambda_{QCD}$), it is also significantly larger than $u$ and $d$. Hence, the strangeness ($S$) of a system becomes an important identifier for comparing, in theories and experiments, phenomena which reflect the complex mechanisms of quantum chromodynamics (QCD) in a hadronic system. 
Therefore, it is essential to understand not only $S=-1$ systems containing one strange quark, but also $S=-2$ systems, particularly $\Lambda$$\Lambda$ and $\Xi N$ interactions, through studies of double hypernuclei.
$H$-dibaryons composed of 6 quarks ($uuddss$) are predicted to exist in $S=-2$ systems~\cite{Jaffe:1976yi}, and whether they do or not is particularly significant from the perspective of hadron physics.
In addition, construction of baryon--baryon interactions can serve as an important clue for predicting the behavior of the equation of state of nuclear matter under high-density conditions where hyperons are expected to appear.
In particular, such studies are indispensable to understand how baryons exist inside neutron stars~\cite{SchaffnerBielich:2008kb}. 

Regarding $S=-1$ systems ({\it i.e.}, $\Lambda N$ or $\Sigma N$ interactions) progress is being made through studies of single $\Lambda$ hypernuclei containing one $\Lambda$ hyperon in the nucleus~\cite{Davis:2005mb,Hashimoto:2006aw}, and studies of $\Sigma$ atoms and $\Sigma$ hypernuclei~\cite{Gal:2016boi}. 
On the other hand, progress has been more difficult to explore $S=-2$ systems, {\it i.e.}, the $\Lambda$$\Lambda$ and the $\Xi N$ interactions. 
The reason why is because, in order to bind two strange quarks to a nucleus, it is necessary to insert a $\Xi^-$ hyperon, having a shorter lifetime, into a nucleus, with a smaller production cross section than for a $\Lambda$ hyperon. 
In spite of such difficulties, a sequential decay event of a double $\Lambda$ hypernucleus in which the nucleus contains two $\Lambda$ hyperons was discovered in 1963 using nuclear emulsion irradiated with $K^-$ beams; the double $\Lambda$ hypernucleus exhibited a strongly-attractive $\Lambda$$\Lambda$ interaction energy of $\Delta B_{\Lambda\Lambda}$=4.5 $\pm$ 0.4~MeV or 3.2 $\pm$ 0.6~MeV~\cite{Danysz:1963zza}.
By 1979, a several events were also reported in nuclear emulsion, where two single $\Lambda$ hypernuclei were emitted from a reaction point between a $K^-$ meson and a nucleus. 
The binding energy of the $\Xi^-$ hyperon in the system of the $\Xi^-$ and the nucleus ($B_{\Xi^-}$) was obtained in a few cases from the mass of the $\Xi$-nucleus system measured as the energy sum of the two single $\Lambda$ hypernuclei~\cite{Mondal:1979hp}.
Based on these values, the Woods-Saxon (WS) potential depth ($V_0^\Xi$) for the $\Xi^-$ hyperon in the nucleus was estimated to be 21--24~MeV~\cite{Dover:1982ng}.

In order to reliably and efficiently absorb two strange quarks into a nucleus, a new experimental technique, the ``emulsion-counter hybrid method'', has been used since 1989.
$\Xi^-$ hyperons are identified with electronic detectors through a quasi-free reaction ``$p$''($K^-$, $K^+$)$\Xi^-$ between a $K^-$ beam and a proton in the target material.
If a $\Xi^-$ hyperon slows down and stops inside the material, then a $\Xi^-$ atom is produced. 
A $\Xi$ hypernucleus can be formed when the $\Xi^-$ hyperon in the $\Xi^-$ atom is bound in a deeper atomic level.
The $\Xi^-$ hyperon in a deep atomic orbit or in a $\Xi$ hypernuclear state reacts promptly with a proton in the nucleus, and two $\Lambda$ hyperons and nuclear fragments are produced due to the reaction $\Xi^-$ + p $\to$ $\Lambda$ + $\Lambda$ + 28~MeV. 
If a single nuclear fragment captures two $\Lambda$ hyperons, a double $\Lambda$ hypernucleus results.
On the other hand, if each of the two nuclear fragments captures one $\Lambda$ hyperon, a twin single-$\Lambda$ hypernucleus (twin hypernucleus) is produced.

The $\Lambda$ hyperons in a double $\Lambda$ hypernucleus or single $\Lambda$ hypernuclei can undergo mesonic weak decay, in which a $\pi$ meson is emitted, or non-mesonic weak decay, in which a $\pi$ meson is not emitted. 
If the double $\Lambda$ hypernuclear mass is reconstructed from this formation and decay process, then the $\Lambda$$\Lambda$ interaction energy ($\Delta B_{\Lambda\Lambda}$) can be obtained through the binding energy of the two $\Lambda$ hyperons ($B_{\Lambda\Lambda}$).
If the mass of the $\Xi^-$ atom or the $\Xi$ hypernucleus is reconstructed through the binding energy of $\Lambda$ in the nucleus ($B_\Lambda$), then the $\Xi N$ interaction strength can be obtained using its binding energy $B_{\Xi^-}$. 
A $\Xi^-$ atom and a $\Xi$ hypernucleus can be distinguished by the magnitude of $B_{\Xi^-}$.

In the E176 experiment, which started in 1989, approximately 80 events of $\Xi^-$ hyperons stopped in the emulsion were observed~\cite{Aoki:1998sv}. 
One event among these was identified as a double $\Lambda$ hypernucleus, but two interpretations were allowed: \Nuc{Be}{10}{\Lambda\Lambda} and \Nuc{B}{13}{\Lambda\Lambda}~\cite{Aoki:1991ip}. 
Three events of a twin hypernucleus were detected, and in two of them, a $\Xi^-$ hyperon was absorbed into \Nuc{C}{12}{}~\cite{Aoki:2009pvs,10.1143/ptp/89.2.493,Aoki:1995za}. 
When excited states of the daughter single $\Lambda$ hypernuclei are taken into account, a number of possibilities remained for $B_{\Xi^-}$.
The obtained $B_{\Xi^-}$ value probably corresponds to a deeper level than the atomic $3D$ level, suggesting the existence of $\Xi$ hypernuclei in the Coulomb-assisted nuclear $1p$ state. 
If the $1p$ state is assumed, $V_0^\Xi$=~16~MeV is obtained~\cite{yamamoto1994formation}, raising questions about the past value of $V_0^\Xi$=21--24~MeV. 
The KEK E224~\cite{Fukuda:1998bi} and the BNL E885~\cite{Khaustov:1999bz} experiments measured missing-mass spectra of $\Xi$ hypernuclei with a ($K^-$, $K^+$) reaction on carbon target.
The BNL E885 is in reasonable agreement with the calculated spectrum assuming that $V_0^\Xi$ is 14~MeV, which is consistent with the E176 result.
However, the resolutions of the E224 and the E885 experiments were insufficient to separate the peak structure in the bound region of $\Xi$ hypernuclei.
After that, in the E373 experiment~\cite{Ichikawa:1998yk} which aimed to detect 10 times more double hypernuclei than the E176 experiment, and the J-PARC E07 experiment~\cite{ProposalE07} which aimed to detect 10 times the events of E373, approximately 650 events\cite{Theint:2019wkg} and 3300 events (by September 2019) of $\Xi^-$ hyperons stopped in the emulsion were detected, respectively.

A double $\Lambda$ hypernucleus detected in the E373 experiment, called the NAGARA event, was identified as \Nuc{He}{\,\,6}{\Lambda\Lambda} with no uncertainty in the interpretation~\cite{Takahashi:2001nm}. 
The results, $B_{\Lambda\Lambda}$=6.91 $\pm$ 0.16~MeV and $\Delta B_{\Lambda\Lambda}$=0.67 $\pm$ 0.17~MeV obtained assuming $B_{\Xi^-}$=0.13~MeV for the $3D$ level~\cite{Ahn:2013poa} when a $\Xi^-$ hyperon was absorbed by \Nuc{C}{12}{}, superseded the past value of the $\Lambda\Lambda$ interaction energy obtained from the old double $\Lambda$ hypernuclear event in the 1960s.
In the E07 experiment, 14 events of double $\Lambda$ hypernuclei having a clear topology were detected, and of these, the MINO event was already reported~\cite{Ekawa:2018oqt}.

On the other hand, a new analysis method, the so-called ``overall scanning method'', was developed~\cite{Yoshida:2017oww}.
The method searches the entire volume of emulsion sheets for three branch tracks characteristic of double hypernuclear events with image processing technology.
The KISO event, a twin hypernuclear event detected in the E373 emulsion with the overall scanning method, was uniquely identified as a decay of $\Xi^-$ + \Nuc{N}{14}{} $\to$ \Nuc{Be}{10}{\Lambda}+\Nuc{He}{5}{\Lambda}~\cite{Nakazawa:2015joa,Hiyama:2018lgs}.
The $B_{\Xi^-}$ was 3.87 $\pm$ 0.21 or 1.03 $\pm$ 0.18~MeV, and this was significantly deeper than the value of $B_{\Xi^-}$=0.17~MeV for the atomic $3D$ state of the $\Xi^-$--\Nuc{N}{14}{} system. 
Because the KISO event clearly showed that a $\Xi$ hypernucleus exists, the $\Xi N$ interaction is found to be attractive for the first time.
Furthermore, the IBUKI event~\cite{Hayakawa:2020oam}, a twin hypernuclear event detected in the E07 emulsion, was uniquely identified to have the same decay mode as the KISO event.
The $B_{\Xi^-}$ of the IBUKI event was measured to be 1.27 $\pm$ 0.21~MeV and is interpreted as a $\Xi$ hypernucleus with a $\Xi^-$ hyperon in a nuclear $1p$ state.

Experiments to study $\Xi N$ interaction using electronic detectors have also been preformed.
The J-PARC E05 experiment was conducted to observe missing-mass spectra from the \Nuc{C}{12}{}($K^-$, $K^+$)\Nuc{Be}{12}{\Xi} reaction with a higher resolution than that in the E224 and the E885 experiments.
The J-PARC E70 experiment with even higher resolution is also planned.
An energy shift and width of X-rays from $\Xi^-$ atoms are expected to provide clear information on $\Xi N$ interaction~\cite{Batty:1998fn}.
The J-PARC E07 made the first attempt to measure coincident X-rays, and the J-PARC E03, to detect X-rays from $\Xi^-$-Fe atoms, is also being performed.
Separately, the ALICE experiment observed an attractive p-$\Xi^-$ strong interaction in $p$-$\Xi^-$ correlation from $p$-Pb collisions with $\sqrt{s_{NN}}$ = 5.02~TeV.

Theoretical studies have been developed alongside of new experimental data.
Based on one boson exchange (OBE) potentials with flavor SU(3) symmetry using rich $NN$ scattering data and sparse $\Lambda N$ and $\Sigma N$ scattering data, the Nijmegen group has developed various baryon-baryon interaction models~\cite{Nagels:1976xq,Nagels:1978sc,Maessen:1989sx,Rijken:1998yy}.
Most of them predict a repulsive $\Xi N$ interaction, while a recent model called ESC16 presented an attractive $\Xi N$ interaction~\cite{Nagels:2015lfa}.
The Ehime model was developed based on a OBE potential with a $\sigma$ meson added to the scalar nonet mesons which couples $\Xi N$, which 
reproduced the experimental value of $B_{\Xi^-}$ $\sim$0.6~MeV\footnote{Though the $B_{\Xi^-}$ value of $\Xi^-$--\Nuc{C}{12}{} system was revised to $\sim$0.8~MeV in Ref.~\cite{Hiyama:2018lgs}, the Ehime model calculation used the old value of $\sim$0.6~MeV.} for $\Xi^-$--\Nuc{C}{12}{} in nuclear $1p$ state from the E176.
Using the Ehime model, energy levels of atomic states for $\Xi^-$--\Nuc{C}{12}{}, $\Xi^-$--\Nuc{N}{14}{} and $\Xi^-$--\Nuc{O}{16}{} were presented in~\cite{Yamaguchi:2001ip}.
The $B_{\Xi^-}$ values were also calculated based on Relativistic Mean Field (RMF) theory and Skyrme-Hartree-Fock (SHF) theory~\cite{Sun:2016tuf} using the $B_{\Xi^-}$ value of the KISO event. 
A calculation based on Quark Meson Coupling (QMC) model was conducted without fitting to the experimental results, and then predicted the $B_{\Xi^-}$ for the $1s$ and the $1p$ states of $\Xi^-$--\Nuc{N}{14}{}~\cite{Guichon:2008zz,Shyam:2019laf}.
The chiral effective field theory (EFT), which incorporates explicitly the scales and symmetries of Quantum Chromodynamics, is a new powerful tool for understanding hadronic interactions. The recent chiral EFT study presented properties of the $\Xi^-$ hyperon in the nuclear medium at the next-to-leading order level \cite{Polinder:2007mp, Kohno:2019oyw}.
The HAL QCD group has started to provide baryon-baryon interactions with lattice QCD simulation~\cite{Ishii:2006ec}, which is independent of phenomenology.
Recently, a lattice QCD calculation near the physical point ($m_{\pi} \simeq$ 146~MeV, $m_{K} \simeq$ 525~MeV) presented a $\Xi N$ potential~\cite{Sasaki:2019qnh}.

By September 2020, 10 events of double hypernuclei (double $\Lambda$ hypernuclei or twin single-$\Lambda$ hypernuclei) were observed in the E373 emulsion and 33 events in the E07 emulsion. 
Of those, there were 3 and 13 events, respectively, of twin hypernuclear events having a clear topology. 
This paper reports on one event (E373-T3) in the E373 experiment and three events (E07-T007, E07-T010, E07-T011) in the E07 experiment whose analyses were recently completed. 
It is interpreted that two events among these exhibit $\Xi^-$ bound states much deeper than the KISO and the IBUKI events.
The present result has a decisive impact on unveiling the nature of the $\Xi$ hypernucleus from 30 years' study of double hypernuclei using nuclear emulsion since the E176 experiment.

\section{Experimental methods}
\begin{figure}[!ht]
    \centering\includegraphics[width=5.0in]{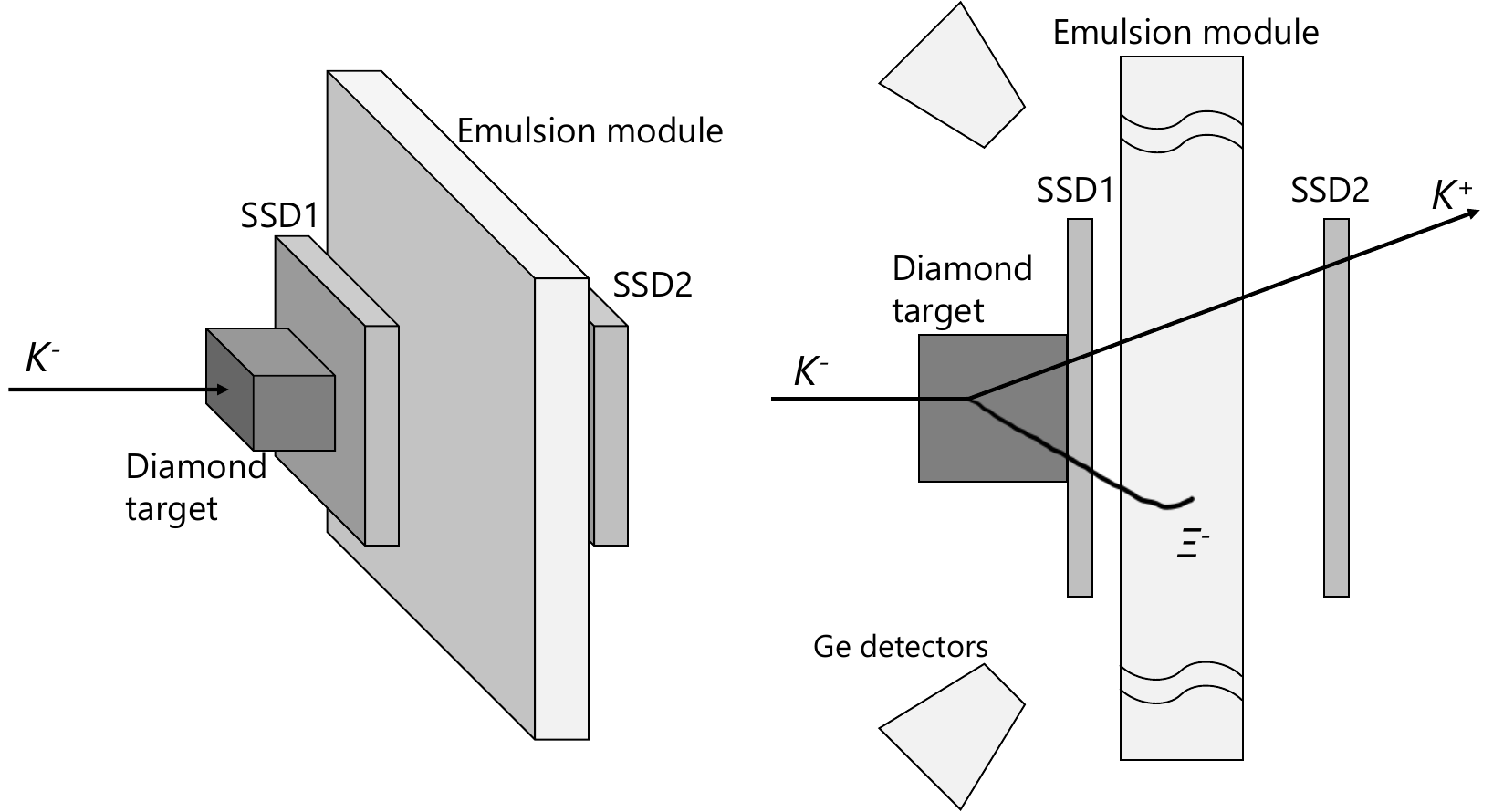}
    \caption{\label{fig:E07}Schematic drawing of experimental setup of the E07 experiment around the emulsion module.}
\end{figure}

The E373 and the E07 experiments, with the goal of detecting double hypernuclei, were conducted on the $K^-$ beamlines at KEK proton synchrotron (KEK-PS) in 1998--2000 and at J-PARC in 2016--2017, respectively. 
The quasi-free p($K^-$, $K^+$)$\Xi^-$ reactions in a diamond target with a thickness of 10~g/cm$^2$ were selected by using a beamline magnetic spectrometer for the incoming $K^-$ and a KURAMA magnetic spectrometer for the outgoing $K^+$. 
The $\Xi^-$ productions were identified by the missing mass technique and the track of the $\Xi^-$ hyperons were detected with a scintillating microfiber-bundle detector (SciFi-Bundle) in the E373 experiment and with a silicon strip detector (SSD) in the E07 experiment as shown in Fig.~\ref{fig:E07}. 
In the E07 experiment, an array of germanium (Ge) detectors, called Hyperball-X was mounted around the emulsion module to detect the $\Xi^-$-atomic X-rays~\cite{Fujita:2020}.

Emulsion gel produced by the FUJIFILM Corporation was applied to a plastic film and dried, resulting in an emulsion sheet.
The atomic composition of the dried emulsion layer used in the E07 experiment is shown in Table~\ref{tab:Composition}.
This composition was predicted from the raw materials and the production process of the emulsion gel.
The typical density of the dried emulsion layer was 3.53~g/cm$^3$.
Thick emulsion sheets to stop the $\Xi^-$ hyperons and to observe the $\Xi^-$-stop events were formed from two emulsion layers with the thickness of 480~$\mu$m on both sides of a 40~$\mu$m polystyrene film.
Thin emulsion sheets to measure the angle of penetrating tracks were formed from two 100~$\mu$m emulsion layers on a 180~$\mu$m polystyrene film as a base.
An emulsion module consists of these 13 sheets where the 2 thin sheets sandwiched the 11 thick sheets.
The design concept of an emulsion module in the E373 experiment is the same as that of the E07.

Table~\ref{tab:Comparison} shows the numerical values for the scale of the E373 and the E07 experiments.
The purity of the $K^-$ beam in the E07 experiment was 3.3 times higher than that in the E373, and the weight of the emulsion gel in the E07 was 2.6 times larger than that in the E373. 
The numbers of $K^-$ were 1.4$\times$10$^{10}$ and 1.13$\times$10$^{11}$ for the E373 and the E07 experiments, respectively.
Assuming that detection efficiency of $\Xi^-$-stop events is the same in the two experiments, then $\Xi^-$-stops in the E07 are expected to be approximately 8 times more events than those in the E373. 
More details of the E373 and the E07 experiments are given in Ref.~\cite{Ichikawa:2001hh,Ahn:2013poa} and Ref.~\cite{Hayakawa:2019dth,Ekawa:2020fya}, respectively. 

\begin{table}[!ht]
    \centering
    \caption{\label{tab:Composition}Atomic composition of the dried emulsion layer of nuclear emulsion in the E07 experiment.}
    \begin{tabular}{rrrrrrrr}
    \hline
                      &H      & C     & N    & O     & Br    & Ag   & I    \\
    \hline
    Mass ratio [\%]   &  1.42 & 9.27  & 3.13 & 6.54  & 33.17 & 45.52& 0.94 \\
    Atomic ratio [\%] & 38.56 & 21.10 & 6.10 & 11.17 & 11.34 & 11.53& 0.20 \\
    \hline
    \end{tabular}
\end{table}

\begin{table}[!ht]
    \centering
    \caption{\label{tab:Comparison}Total weight of the emulsion gel to make the emulsion sheets, the purity of $K^-$ in the beam, and the amount of $K^-$ irradiated in the E373 and the E07 experiments.}
    \begin{tabular}{ccccc}
    \hline
    Experiment    & Emulsion gel & $K^-$ purity & Number of $K^-$\\
                  & [10$^3$~kg] & [\%] & [$\times10^9$]\\
    \hline
    KEK-PS E373  & 0.8     & 25    & 14\\
    J-PARC E07 & 2.1     & 82   & 113\\
    \hline
    \end{tabular}
\end{table}

The candidates of $\Xi^-$ tracks were detected, and the position and angles were estimated, by the electronic detectors. 
The pertinent locations on the thin emulsion sheet in the most upstream of the emulsion module were read with dedicated optical microscopes, and then tracks matching the predicted position and angle were picked up as the $\Xi^-$ hyperon candidates.
A track-following procedure, to locate where $\Xi^-$ hyperons lose the kinetic energy and stop in the emulsion, was carried out using a manual technique in the E373 and an automated scanning technique developed in the E07~\cite{MyintKyawSoe:2017gpq}.
Twin hypernuclear events were identified by the topology around the $\Xi^-$ stop, {\it i.e.}, two single $\Lambda$ hypernuclear tracks emitted from the $\Xi^-$ stop and each single $\Lambda$ hypernucleus decaying into other particles. 

The ranges and angles of all charged particles were precisely measured using images acquired with the microscope. 
The calibration of the range and the kinetic energy ($KE$) was measured using about 100 tracks of alpha particles with a $KE$ of 8.784~MeV from \Nuc{Po}{212}{} of the thorium series in the emulsion.
The emulsion layers shrink due to the photographic developing process, and thus the range was found by appropriately correcting for the shrinkage, and the density was found as a parameter when converting the range to the $KE$. 
Table~\ref{tab:AlphaRange} shows the density calculated from the range of alpha particles in emulsion sheets where twin hypernuclear events were found. 

\begin{table}[!ht]
    \centering
    \caption{\label{tab:AlphaRange}Density of the emulsion layer and the range of alpha particles from \Nuc{Po}{212}{} for each twin hypernuclear event.}
    \begin{tabular}{cS[table-format=2.2,table-figures-uncertainty=1]S[table-format=1.3,table-figures-uncertainty=1]c}
    \hline
    {Event}    & {Range [$\mu$m]} & {Density [g/cm$^3$]} & \# of alphas\\
    \hline
    E373-T3  & 48.80+-0.67     & 3.71 +-0.08    & 82\\
    E07-T007 & 50.05+-0.15     & 3.567+-0.017   & 94\\
    E07-T010 & 48.79+-0.10     & 3.714+-0.012   &165\\
    E07-T011 & 50.63+-0.10     & 3.502+-0.011   &147 \\
    \hline
    \end{tabular}
\end{table}

For the following analysis, we assumed light elements, \Nuc{C}{12}{}, \Nuc{C}{13}{}, \Nuc{N}{14}{}, and \Nuc{O}{16}{} as the nuclei absorbing a $\Xi^-$ hyperon because their atomic ratios are over 0.1\% in the emulsion. 
We excluded heavy elements, bromine, silver, and iodine, although the probability of $\Xi^-$ hyperons being captured by the heavy elements is about half that of all elements. 
No visible fragment of a single $\Lambda$ hypernucleus was found associated with any of the 30 and 237 $\sigma$-stops ({\it i.e.}, $\Xi^-$ hyperons stopped in the emulsion with visible fragments at its stopping point~\cite{Aoki:1998sv}), as seen in capture by heavy elements in the E176~\cite{Aoki:2009pvs} and the E373~\cite{Theint:2019wkg}, respectively.

For the emitted charged particles ({\it i.e.}, visible tracks), all decay modes were examined, assuming all possible single $\Lambda$ hypernuclei, ordinary nuclei and a $\pi^-$ meson. 
Letting the initial state rest mass be $M_{\Xi^-}+M_X$, where $X$ is a nucleus absorbing a $\Xi^-$ hyperon, and the final state invariant mass be $m_0$, $B_{\Xi^-}$ is given by $B_{\Xi^-}=M_{\Xi^-}+M_X-m_0$.
The $B_\Lambda$ of single $\Lambda$ hypernuclei to calculate the rest mass was obtained with reference to \cite{Juric:1973zq,Hashimoto:2006aw,Bertini:1981zx,May:1981er,Dluzewski:1988ye,Davis:1986kg,Cusanno:2008xx,Botta:2012xi,Gogami:2015tvu}.
The $B_\Lambda$ of possible single $\Lambda$ hypernuclei, not yet measured experimentally, were obtained by linear approximation from known $B_\Lambda$ with the same atomic number.
Major mass parameters used in the present analyses are listed in \ref{app:mass}.

The conservation laws of momentum and mass–energy were examined assuming that no neutrons are emitted from $\Xi^-$-stop (no-neutron mode).
If the total momentum ($\vec{p}_{tot}$) composed of visible particles was not 0 or the inequality $B_{\Xi^-}\geq0$ was not satisfied, even taking account of a 3$\sigma$ error, the decay modes were excluded from the analyses.
On the other hand, in decay modes emitting neutrons (with-neutron mode), the $\vec{p}_{tot}$ including the neutrons is always 0, since the momentum of the neutrons was taken to be opposite of the sum of visible momenta.
Therefore, only the conservation law of the mass–energy was used.
When examining the single $\Lambda$ hypernuclear decay modes, it was assumed in addition that a $\pi^0$ was emitted.

The error in the momentum of a visible particle was deduced from the range measurement error and the range straggling estimated from the range. 
The optimal momenta and angles of visible particles in the no-neutron mode were found with Lagrange's method of undetermined multipliers~\cite{Avery:1991}, taking $\vec{p}_{tot}=0$ as the constraint condition.
The $B_{\Xi^-}$ and the error were obtained using the optimal momenta and the angles.
On the other hand, in the with-neutron mode, since the $\vec{p}_{tot}$ including neutrons always becomes 0, the Lagrange's method of undetermined multipliers cannot be applied. 
Thus, the $B_{\Xi^-}$ error in the with-neutron mode was deduced through a simple error propagation. 
The error in the $KE$ of the neutrons was propagated from the error in the momentum of the neutron. 
The error in the $B_{\Xi^-}$ value was defined as the quadratic sum of the $KE$ errors of the all emitted particles. 
Finally, errors in rest mass of the single $\Lambda$ hypernuclei and the $\Xi^-$ hyperon were taken into account.

\clearpage
\section{Twin hypernuclear events}
\subsection{E373-T3 (KINKA) event}
\begin{figure}[!ht]
    \centering\includegraphics[width=4.5in]{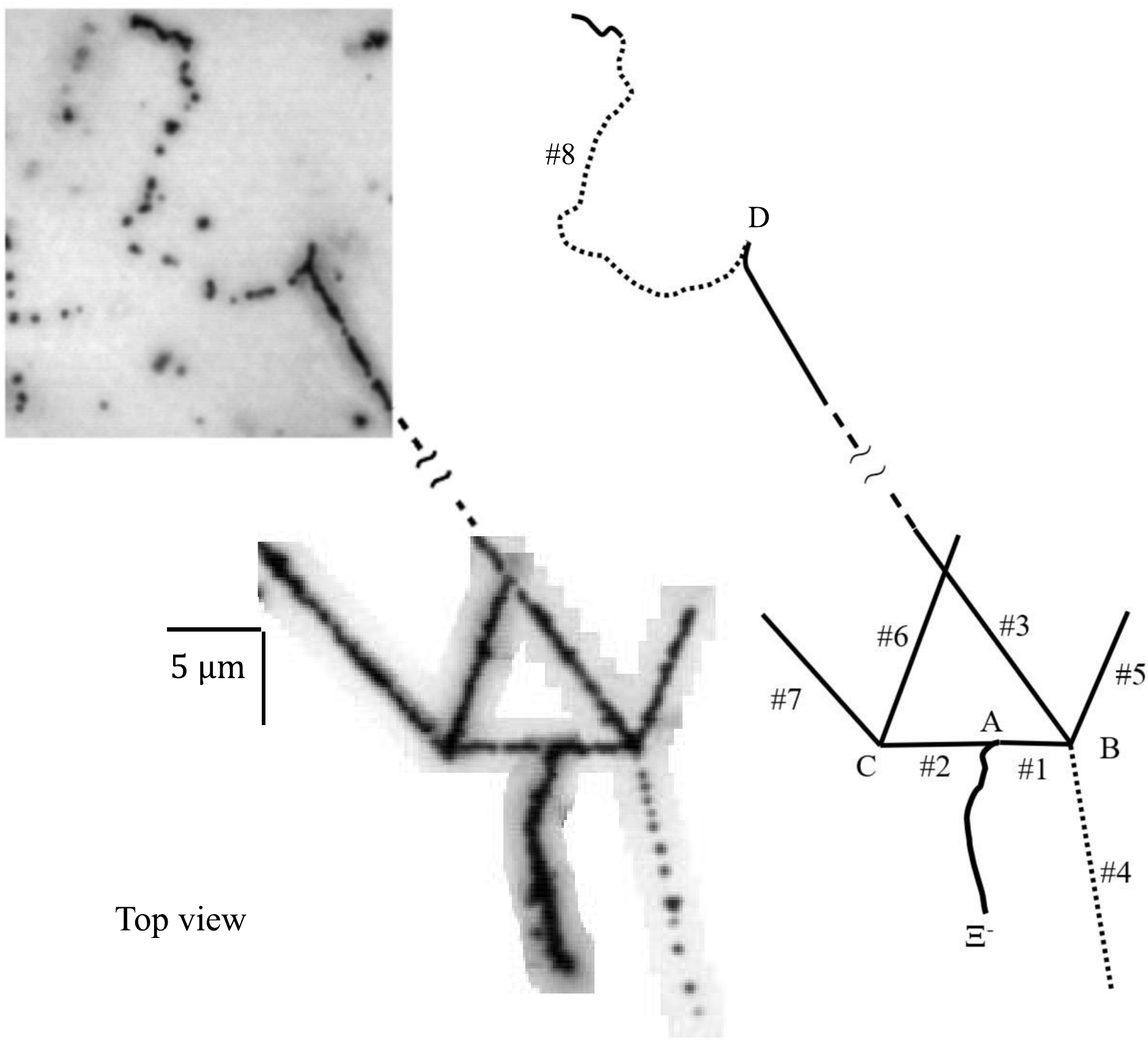}
    \caption{A superimposed photo and a schematic drawing of the E373-T3 (KINKA) event from the top view of the emulsion sheet.}
    \label{fig:T3_vertical}
\end{figure}

A microscope photo and a schematic drawing for the E373-T3 event are shown in Fig.~\ref{fig:T3_vertical}. 
The topology of a twin hypernucleus is shown at vertex A, where two charged particles are emitted, and those particles respectively decay at vertices B and C. 

\begin{figure}[!ht]
    \centering\includegraphics[width=3.5in]{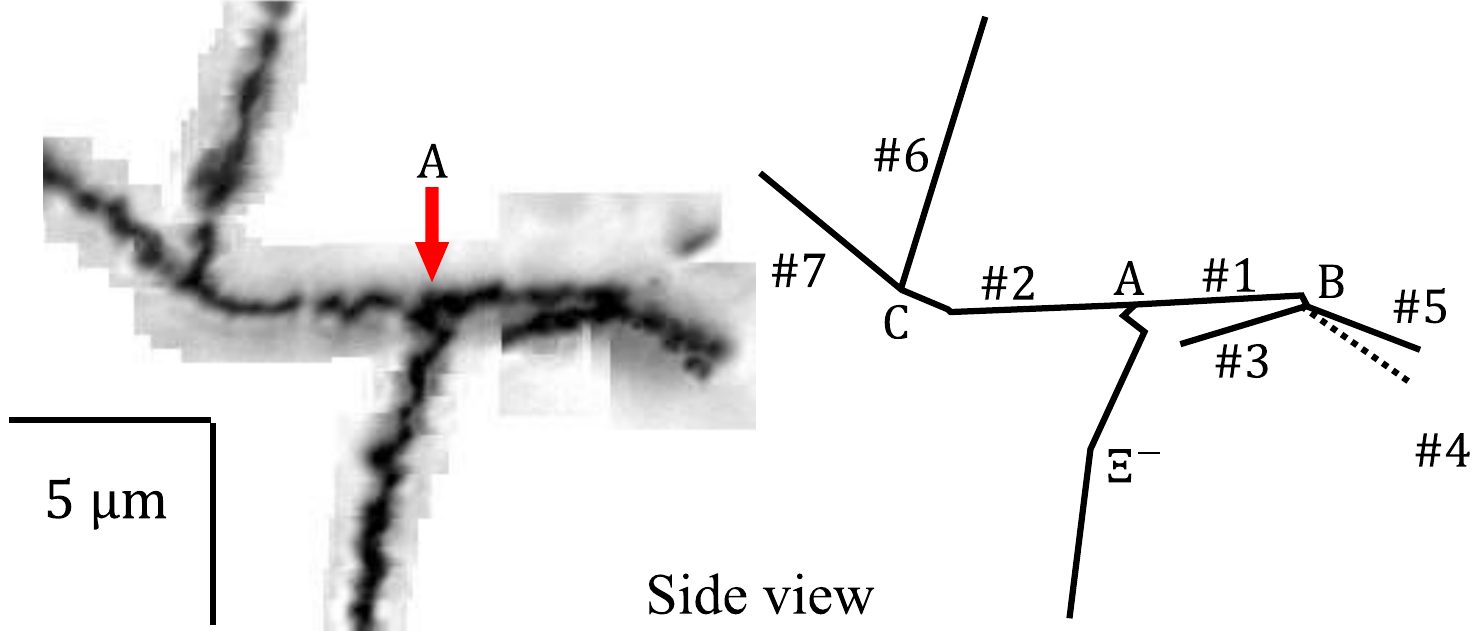}
    \caption{A superimposed photo and a schematic drawing of the E373-T3 (KINKA) event from the side view of the emulsion sheet.}
    \label{fig:T3_horizontal}
\end{figure}

The stopping point of the $\Xi^-$ hyperon, vertex A in the microscope photo of Fig.~\ref{fig:T3_vertical}, could not be precisely measured in the usual way, from the top view, because the tracks of the two single $\Lambda$ hypernuclei overlapped with the $\Xi^-$ hyperon track in the optical axis direction.
Therefore, the method of viewing the event from the side was used, which has also been done in past event analysis~\cite{Ichikawa:2001hh}.
Figure~\ref{fig:T3_horizontal} shows a microscope photo when the thinly sliced emulsion sheet was observed from the side. 
When observed from this direction, the $\Xi^-$ hyperon track is separated from the single $\Lambda$ hypernuclear tracks, and therefore it was found that the $\Xi^-$-stopping point is located at the tip of the arrow in Fig.~\ref{fig:T3_horizontal}.

Table~\ref{tab:T3RTP} shows the ranges and angles of all charged particles in E373-T3. 
The $\theta$ and $\phi$ are the zenith and azimuth angles, respectively, when the emulsion sheet is viewed as a fundamental plane.
The ranges of tracks \#1 and \#2 were measured with the microscope photo from the side view. 
Other values, including the sum of the ranges of tracks \#1 and \#2 were measured with the microscope photo from the top view.

Of all the decay modes examined at vertex A, Table~\ref{tab:T3_PointA} shows the modes satisfying the condition $B_{\Xi^-}\geq0$~MeV and $\vec{p}_{tot}=0$ within the $3\sigma$ error. 
Four nuclides of candidates (\Nuc{He}{5}{\Lambda}, \Nuc{Li}{7}{\Lambda}, \Nuc{Li}{8}{\Lambda}, and \Nuc{Be}{9}{\Lambda}) are possible interpretation for tracks \#1 and \#2.

Tracks \#3, \#4, and \#5 were emitted from vertex B. 
Track \#4 penetrated the most upstream emulsion sheet, and thus only a lower limit range was obtained. 
The meandering track (\#8) from the stopping point (D) of track \#3 in Fig.~\ref{fig:T3_vertical} was observed. 
The track with the characteristics is a low-energy electron. 
Therefore, it was deduced that track \#3 was \Nuc{He}{6}{}, \Nuc{He}{8}{}, or \Nuc{Li}{9}{}, nuclides which undergo $\beta$ decay.
Taking into account the number of nucleons in tracks \#1 and \#3, \Nuc{Be}{9}{\Lambda} is only acceptable for the nucleus in track \#1.
The decay mode at vertex B was uniquely identified to be \Nuc{Be}{9}{\Lambda} (\#1) $\to$ \Nuc{He}{6}{} (\#3) + p (\#4) + p (\#5) + n with the Q-value of 142.2~MeV and the $KE_{total} >$81.4~MeV.
The particle charge in track \#3 is likely to be $+2$, see Ref.~\cite{Kinbara:2019kyx}, supporting the present interpretation of \Nuc{He}{6}{}.

Tracks \#6 and \#7 were observed from vertex C. 
Three decay modes were accepted at vertex C as shown in Table~\ref{tab:T3_C}, and these were all solutions taking the \Nuc{He}{5}{\Lambda} as the single $\Lambda$ hypernucleus in track \#2. 
Thus the single $\Lambda$ hypernucleus in track \#2 is found to be \Nuc{He}{5}{\Lambda}.
Since the nuclides of tracks \#1 and \#2 were identified, the decay mode at vertex A was determined uniquely, {\it i.e.}, $\Xi^-$ + \Nuc{N}{14}{} $\to$ \Nuc{Be}{9}{\Lambda} (\#1) + \Nuc{He}{5}{\Lambda} (\#2)+ n. 

In the energy range of twin hypernuclear production and decay, an approximation of non-relativistic kinematics is possible.
Hence, the error in the $KE$ is proportional to the momentum and its error. 
As the momentum of the neutron emitted from vertex A is 103.8 $\pm$ 6.8~MeV/$c$, the $KE$ of the neutron was estimated to be 5.71 $\pm$ 0.75~MeV.
The sum of the $KE$ of the visible particles at vertex A was estimated to be 4.87 $\pm$ 0.18~MeV.
Therefore, the $B_{\Xi^-}$ of the E373-T3 was found to be 8.00 $\pm$ 0.77~MeV if the daughter nucleus, \Nuc{Be}{9}{\Lambda}, is in the ground state.
If the \Nuc{Be}{9}{\Lambda} is in the excited state of 3.04 $\pm$ 0.02~MeV~\cite{Akikawa:2002tm} described in \ref{app:mass}, the $B_{\Xi^-}$ is 4.96 $\pm$ 0.77~MeV. 
We named E373-T3 KINKA, which is a mountain in Gifu prefecture, Japan.

\begin{table}[!ht]
    \centering
    \caption{\label{tab:T3RTP}Ranges and angles of associated tracks of E373-T3 (KINKA) event.
    The range of track \#1+\#2 was measured in the photo from the top view. The ranges of the tracks \#1 and \#2 were measured in the photo from the side view.}
    \begin{tabular}{llcS[table-format=3.1,table-figures-uncertainty=1]S[table-format=3.1,table-figures-uncertainty=1]l}
        \hline
        Vertex & Track & {Range [$\mu$m]}    & {$\theta$ [deg]} & {$\phi$ [deg]} & Comments             \\
        \hline
    A      & \#1      & 4.34 $\pm$ 0.20        & 95.2 +- 2.3     & 183.4 +- 2.7    & Single $\Lambda$ hypernucleus         \\
           & \#2      & 5.74 $\mp$ 0.20        & 81.2 +- 1.3     & 1.4 +- 1.1      & Single $\Lambda$ hypernucleus         \\
           & \#1+\#2  & 10.08 $\pm$ 0.17       & & &                             \\
    B      & \#3      & 75.39 $\pm$ 0.40       & 79.0 +- 0.4     & 305.2 +- 0.2    & $\beta$-decay at the end point    \\
           & \#4      & {$>$6110}                & 72.8 +- 0.4     & 103.4 +- 0.9    & Passed through the emulsion \\
           & \#5      & 7.79 $\pm$ 0.40        & 88.6 +- 1.2     & 249.3 +- 1.9    &                             \\
    C      & \#6      & 4338.4 $\pm$ 1.6       & 149.6 +- 0.3    & 254.1 +- 0.9    &                             \\
           & \#7      & 1185.3 $\pm$ 0.9       & 122.6 +- 0.2    & 313.4 +- 0.1    &                            \\
        \hline
    \end{tabular}
\end{table}

\begin{table}[!ht]
    \centering
    \caption{Possible decay modes of E373-T3 (KINKA) at vertex A satisfying the conservation laws of momentum and mass-energy.\label{tab:T3_PointA}}
    \begin{tabular}{llllS[table-format=2.2,table-figures-uncertainty=1]l}
    \hline
    Parents    & \#1&\#2&        & {$B_{\Xi^-}$}        \\
    \hline
    $\Xi^-$--\Nuc{C}{12}{} & \Nuc{Li}{7}{\Lambda} & \Nuc{He}{5}{\Lambda} & n  & +0.50 +- 0.29 \\
    $\Xi^-$--\Nuc{C}{13}{} & \Nuc{He}{5}{\Lambda} & \Nuc{Li}{8}{\Lambda} & n  & -0.88 +- 0.60 \\
                            & \Nuc{Li}{8}{\Lambda} & \Nuc{He}{5}{\Lambda} & n  & +3.60 +- 0.35 \\
    $\Xi^-$--\Nuc{N}{14}{} & \Nuc{He}{5}{\Lambda} & \Nuc{Be}{9}{\Lambda} & n  & -0.77 +- 1.07 \\
                            & \Nuc{Li}{8}{\Lambda} & \Nuc{Li}{7}{\Lambda} &    & +2.46 +- 0.16 \\
                            & \Nuc{Be}{9}{\Lambda} & \Nuc{He}{5}{\Lambda} & n  & +8.00 +- 0.77 \\
    $\Xi^-$--\Nuc{O}{16}{} & \Nuc{Be}{9}{\Lambda} & \Nuc{Li}{8}{\Lambda} &    & +4.17 +- 0.20 \\
    \hline
    \end{tabular}
\end{table}

\begin{table}[!ht]
    \centering
    \caption{Possible decay modes, their Q-values and kinetic energies at vertex C of E373-T3 (KINKA).\label{tab:T3_C}}
    \begin{tabular}{llllcl}
    \hline
    \#2 & \#6 & \#7 &         & Q-value [MeV]        & $KE_{total}$ [MeV] \\
    \hline
    \Nuc{He}{5}{\Lambda} & p & p & 3n  & 144.7 & $>$73.3 \\
                         & p & d & 2n  & 146.9 & $>$113.0 \\
                         & d & p & 2n  & 146.9 & $>$133.1 \\
    \hline
    \end{tabular}
\end{table}

\subsection{E07-T007 event}
\begin{figure}[!ht]
    \centering\includegraphics[width=3.5in]{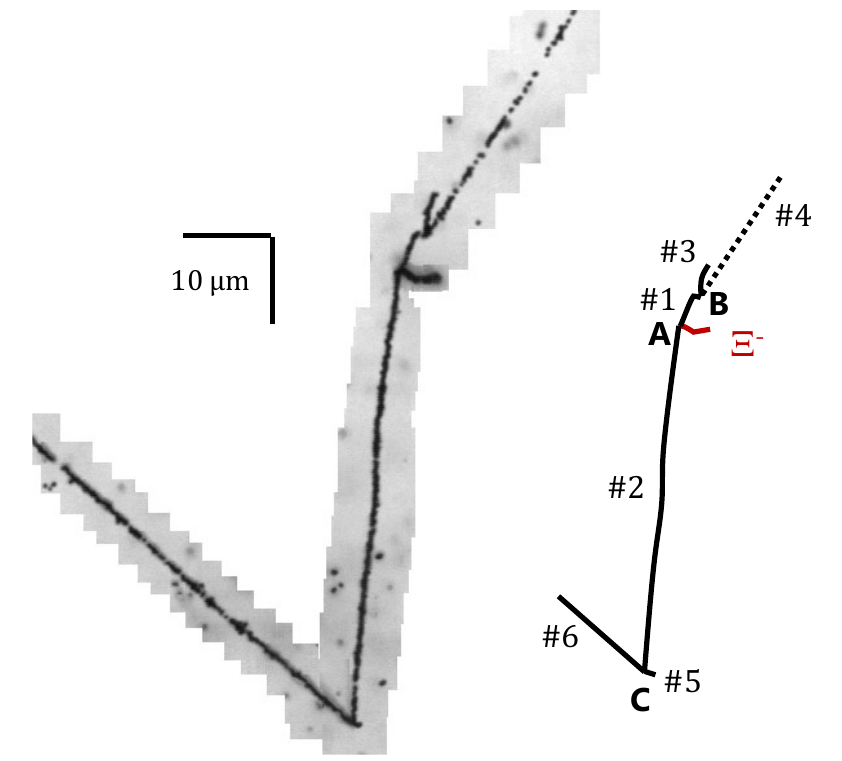}
    \caption{A superimposed photo and a schematic drawing of the E07-T007 event.}
    \label{fig:T007}
\end{figure}

A microscope photo and a schematic drawing for the E07-T007 event are shown in Fig.~\ref{fig:T007}.
Table~\ref{tab:T007RTP} shows the ranges and angles of all charged particles in E07-T007.
The only decay mode satisfying the conservation laws of momentum and mass-energy is $\Xi^-$ + \Nuc{N}{14}{} $\to$ \Nuc{Be}{9}{\Lambda} (\#1) + \Nuc{He}{5}{\Lambda} (\#2) + n. 
All possible decay modes were examined at vertices B and C and many decay modes were accepted.
At vertex B, two decay modes, where track \#1 was \Nuc{Be}{9}{\Lambda}, were accepted as shown in Table~\ref{tab:T007_B}.
Moreover, five decay modes, where track \#2 was \Nuc{He}{5}{\Lambda}, were accepted as shown in Table~\ref{tab:T007_C}.
Therefore, the decay modes at vertices A--C are consistent with each other.
Finally, the $B_{\Xi^-}$ of E07-T007 was found to be $-$1.04 $\pm$ 0.85~MeV, whose daughter \Nuc{Be}{9}{\Lambda} is in the ground state.
Because the $B_{\Xi^-}$, whose \Nuc{Be}{9}{\Lambda} is in the excited state, is largely negative, the case was rejected.

\begin{table}[!ht]
    \centering
    \caption{\label{tab:T007RTP}Ranges and angles of associated tracks of E07-T007 event.}
    \begin{tabular}{llS[table-format=4.2,table-figures-uncertainty=1]S[table-format=3.1,table-figures-uncertainty=1]S[table-format=3.1,table-figures-uncertainty=1]l}
        \hline
        Vertex & Track & {Range [$\mu$m]} & {$\theta$ [deg]} & {$\phi$ [deg]} & Comments             \\
        \hline
    A      & \#1      & 6.45 +-0.2        & 112.1+-1.8      & 294.2+-1.8    & Single $\Lambda$ hypernucleus \\
           & \#2      & 54.76 +-0.2       & 70.9+-1.1       & 96.3+-1.1     & Single $\Lambda$ hypernucleus \\
    B      & \#3      & 5.05 +-0.2        & 68.0+-4.6       & 273.6+-4.6    &                     \\
           & \#4      & 2123.3+-1.1       & 93.0+-0.8       & 304.5+-0.8    &                     \\
    C      & \#5      & 1.0+-0.2          & 112+-11         & 18    +-11    &                     \\
           & \#6      & 336.1+-0.9        & 55.6+-1.2       & 221.2+-1.2    &                     \\
           \hline
        \end{tabular}
\end{table}

\begin{table}[!ht]
    \centering
    \caption{Possible decay modes, their Q-values and kinetic energies at vertex B of E07-T007.\label{tab:T007_B}}
    \begin{tabular}{llllcl}
    \hline
    \#1 & \#3 & \#4 &         & Q-value [MeV]        & $KE_{total}$ [MeV] \\
    \hline
    \Nuc{Be}{9}{\Lambda} & \Nuc{Li}{6}{} & p & 2n  & 144.9 & $>$55.7 \\ 
                         & \Nuc{Li}{6}{} & d &  n  & 147.1 & 142.1 $\pm$ 3.8 \\
    \hline
    \end{tabular}
\end{table}

\begin{table}[!ht]
    \centering
    \caption{Possible decay modes, their Q-values and kinetic energies at vertex C of E07-T007.\label{tab:T007_C}}
    \begin{tabular}{llllcl}
    \hline
    \#2 & \#5 & \#6 &         & Q-value [MeV]        & $KE_{total}$ [MeV] \\
    \hline
    \Nuc{He}{5}{\Lambda} & p & p & 3n       & 144.7 & $>$9.2 \\
                         & p & d & 2n       & 146.9 & $>$17.8 \\
                         & d & p & 2n       & 146.9 & $>$10.0 \\
                         & d & p & $\pi^0$ + 2n & 12.0 & $>$9.8 \\
                         & t & p & $\pi^0$ + n  & 18.2 & $>$11.8 \\
    \hline
    \end{tabular}
\end{table}

\subsection{E07-T010 (IRRAWADDY) event}

\begin{figure}[!ht]
    \centering\includegraphics[width=4.0in]{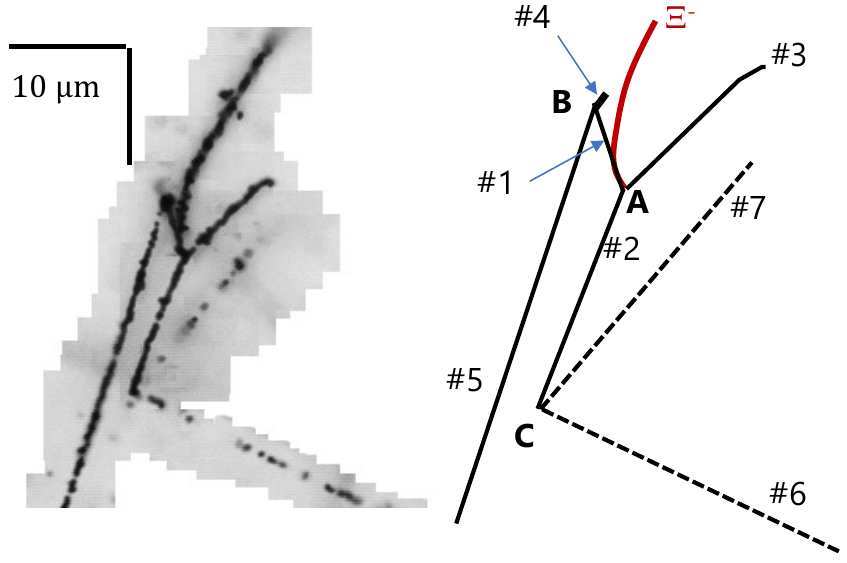}
    \caption{A superimposed photo and a schematic drawing of the E07-T010 (IRRAWADDY) event.}
    \label{fig:T010}
\end{figure}

Figure~\ref{fig:T010} shows a microscope photo and a schematic drawing for the E07-T010 event.
The topology observed has the $\Xi^-$ hyperon stopped at vertex A and emitting three charged particles, two of which decayed at vertices B and C. 
Table~\ref{tab:T010RTP} shows the ranges and angles of all charged particles.
When all decay modes were examined, only the decay mode $\Xi^-$ + \Nuc{N}{14}{} $\to$ \Nuc{He}{5}{\Lambda} (\#1) + \Nuc{He}{5}{\Lambda} (\#2) + \Nuc{He}{4}{} (\#3) + n satisfied the conservation laws.
All possible decay modes were examined at vertices B and C, with 305 and 9 decay modes acceptable at vertices B and C, respectively.
At vertex B, five decay modes, where track \#1 was \Nuc{He}{5}{\Lambda}, were accepted as shown in Table~\ref{tab:T010_B}.
At vertex C, three decay modes, where track \#2 was \Nuc{He}{5}{\Lambda}, were accepted as shown in Table~\ref{tab:T010_C}. 
The decay modes of vertices A--C are consistent with each other.

As the momentum of the neutron emitted from vertex A is 42.2 $\pm$ 5.3 MeV/$c$, the $KE$ of neutron was estimated to be 0.95 $\pm$ 0.24~MeV.
The sum of the $KE$ of visible particles at vertex A was estimated to be 7.88 $\pm$ 0.11~MeV.
Considering the error of the $KE_{total}$ ($\pm$ 0.26~MeV), the rest mass of \Nuc{He}{5}{\Lambda} ($\pm$ 0.02~MeV) and the rest mass of $\Xi^-$ hyperon ($\pm$ 0.07~MeV), the $B_{\Xi^-}$ value became 6.27 $\pm$ 0.27~MeV.
The $B_{\Xi^-}$ is uniquely determined because the daughter nuclei, \Nuc{He}{5}{\Lambda} and \Nuc{He}{4}{}, have no excited states.
We named E07-T010 IRRAWADDY, which is the largest river in Myanmar.

In the KINKA and IRRAWADDY events, the errors in the $KE$ of the neutrons are much larger than the errors in the $KE$ of the visible particles and others.
Therefore, the neutron $KE$ errors dominate the errors in $B_{\Xi^-}$.
The uncertainty in the $B_{\Xi^-}$ of the IRRAWADDY event is small, because the momentum and its error of the neutron is smaller than that in the KINKA event.

\begin{table}[!ht]
    \centering
    \caption{\label{tab:T010RTP}Ranges and angles of associated tracks of E07-T010 (IRRAWADDY) event.}
    \begin{tabular}{llS[table-format=4.2,table-figures-uncertainty=1]S[table-format=3.2,table-figures-uncertainty=1]S[table-format=3.2,table-figures-uncertainty=1]l}
    \hline
    Vertex & Track & {Range [$\mu$m]}    & {$\theta$ [deg]} & {$\phi$ [deg]} & Comments             \\
    \hline
    A      & \#1      & 4.99 +-0.22  & 97.7 +-2.8     & 288.2 +-2.6    & Single $\Lambda$ hypernucleus \\
           & \#2      & 12.31+-0.22  & 88.2 +-1.0     & 68.91 +-0.95    & Single $\Lambda$ hypernucleus \\
           & \#3      & 10.12+-0.24  & 92.7 +-1.5     & 222.2 +-1.2    &                     \\
    B      & \#4      & 2.9 +- 0.2   & 170 +- 4        & 214 +- 4        &                     \\
           & \#5      & 189.7+- 0.6  & 84 +- 1         & 69 +- 1         &                     \\
    C      & \#6      & 2770.5+- 1.3 & 116.7 +- 1.4    & 154.0 +- 0.4    &                     \\
           & \#7      & 8404.2+- 8.5 & 27.4 +- 0.8     & 229.3 +- 1.3    &                     \\
    \hline
    \end{tabular}
\end{table}

\begin{table}[!ht]
    \centering
    \caption{Possible decay modes assuming that the track \#1 is \Nuc{He}{5}{\Lambda}, their Q-values and kinetic energies at vertex B of E07-T010 (IRRAWADDY).\label{tab:T010_B}}
    \begin{tabular}{llllcl}
    \hline
    \#1 & \#4& \#5 &    & Q-value [MeV]        & $KE_{total}$ [MeV] \\
    \hline
    \Nuc{He}{5}{\Lambda} & p & p & 3n  & 144.7 & $>$7.1 \\
                         & p & d & 2n  & 146.9 & $>$13.5 \\
                         & d & p & 2n  & 146.9 & $>$8.0 \\
                         & d & p & $\pi^0$ + 2n  & 12.0 & $>$7.8 \\
                         & t & p & $\pi^0$ + n  & 18.2 & $>$9.8 \\
    \hline
    \end{tabular}
\end{table}

\begin{table}[!ht]
    \centering
    \caption{Possible decay modes assuming that the track \#2 is \Nuc{He}{5}{\Lambda}, their Q-values and kinetic energies at vertex C of E07-T010 (IRRAWADDY).\label{tab:T010_C}}
    \begin{tabular}{llllcl}
    \hline
    \#2 & \#6 & \#7 &    & Q-value [MeV]        & $KE_{total}$ [MeV] \\
    \hline
    \Nuc{He}{5}{\Lambda} & p & p & 3n  & 144.7 & $>$89.0 \\
                         & p & d & 2n  & 146.9 & $>$146.6 \\
                         & d & p & 2n  & 146.9 & $>$120.4 \\
    \hline
    \end{tabular}
\end{table}

\subsection{E07-T011 event}
\begin{figure}[!ht]
    \centering\includegraphics[width=3.5in]{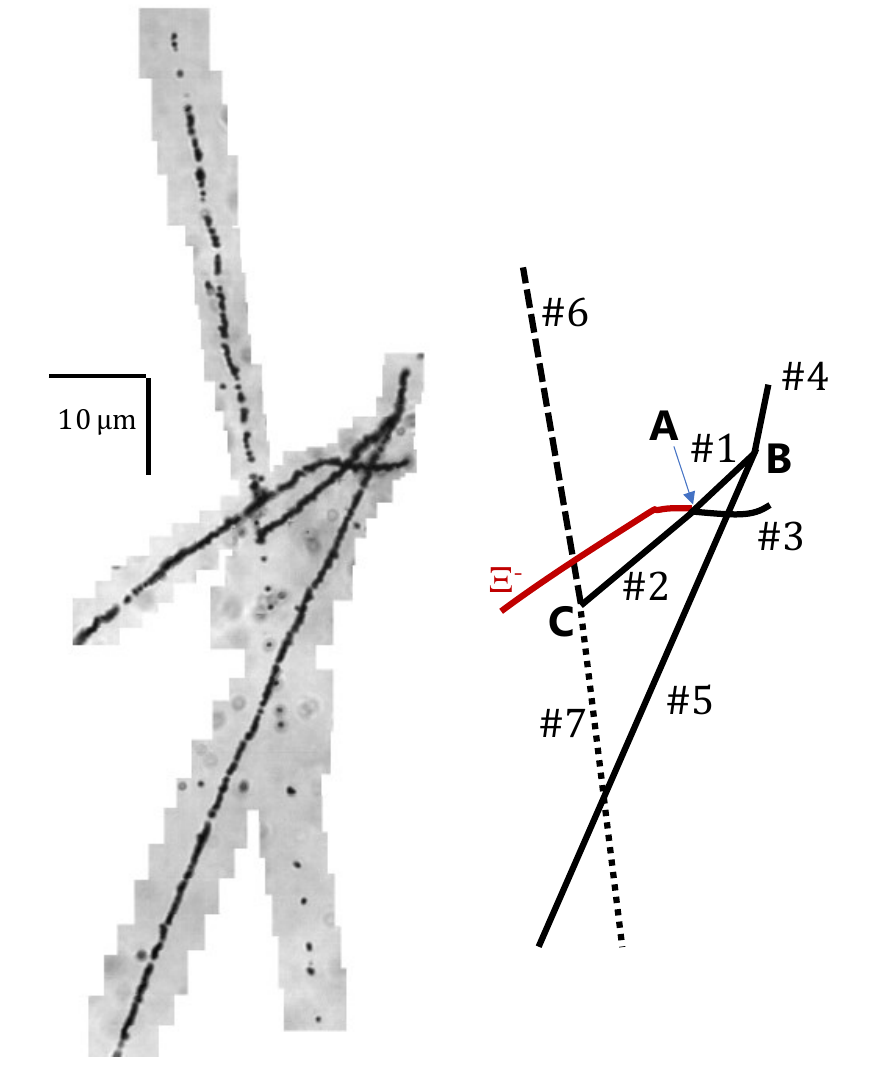}
    \caption{A superimposed photo and a schematic drawing of the E07-T011 event.}
    \label{fig:T011}
\end{figure}

Figure~\ref{fig:T011} shows a microscope photo and a schematic drawing for the E07-T011 event.
Table~\ref{tab:T011RTP} shows the ranges and angles of all charged particles in E07-T011.
Since track \#4 disappeared in the plastic base of the emulsion sheet, 
the shortest range of 5.3 $\mu$m is found if the track stopped at the first boundary between the plastic base and the emulsion layer. 
The longest range of 49.3 $\mu$m is calculated if the track stopped at the second boundary.

The only decay mode at vertex A satisfying the conservation laws was $\Xi^-$ + \Nuc{N}{14}{} $\to$ \Nuc{He}{5}{\Lambda} (\#1) + \Nuc{He}{5}{\Lambda} (\#2) + \Nuc{He}{4}{} (\#3) + n.
All possible decay modes were examined at vertices B and C and many decay modes were accepted.
At vertex B, eight decay modes, where track \#1 was \Nuc{He}{5}{\Lambda}, were accepted in the two types of range of track \#4 as shown in Table~\ref{tab:T011_B}.
At vertex C, five decay modes, where track \#2 was \Nuc{He}{5}{\Lambda}, were accepted as shown in Table~\ref{tab:T011_C}.
Therefore, the decay modes of vertices A--C are consistent with each other.
Finally, the $B_{\Xi^-}$ of E07-T011 was found to be 0.90 $\pm$ 0.62~MeV.
The $B_{\Xi^-}$ value is uniquely determined because there is no excited state for the daughter nuclei.

\begin{table}[!ht]
    \centering
    \caption{\label{tab:T011RTP}Ranges and angles of associated tracks of E07-T011 event.
    The range of track \#4 is the minimum and maximum track length considering the thickness of the plastic base, assuming that the track is straight.}
    \begin{tabular}{llS[table-format=4.2,table-figures-uncertainty=1]S[table-format=3.1,table-figures-uncertainty=1]S[table-format=3.1,table-figures-uncertainty=1]l}
        \hline
        Vertex & Track & {Range [$\mu$m]}& {$\theta$ [deg]}& {$\phi$ [deg]}& Comments             \\
        \hline
    A      & \#1      & 10.58+-0.28      & 131.1+-1.3      & 223.3+-1.5    & Single $\Lambda$ hypernucleus \\
           & \#2      & 12.96+-0.28      & 62.6+-1.4       & 41.7+-1.3     & Single $\Lambda$ hypernucleus \\
           & \#3      & 8.42+-0.22       & 43.9+-1.9       & 175.6+-1.7    &                     \\
    B      & \#4      & {5.3--49.5}      & 116.9+-2.2      & 261.9+-2.2    & Stop in the plastic base    \\
           & \#5      & 113.1+-2.3       & 65.2+-1.2       & 63.7+-1.2     &                     \\
    C      & \#6      & 1998.5+-1.9      & 102.2+-1.2      & 276.3+-1.2    &                     \\
           & \#7      & {$>$10830}       & 72.5+-1.2       & 98.9+-1.2     &                     \\
           \hline
        \end{tabular}
\end{table}

\begin{table}[!ht]
    \centering
    \caption{Possible decay modes, their Q-values and kinetic energies at vertex B of E07-T011.\label{tab:T011_B}}
    \begin{tabular}{llllcll}
    \hline
    \#1 & \#4 & \#5 &         & Q-value [MeV]        & $KE_{total}$ [MeV] & $KE_{total}$ [MeV] \\
        &     &     &         &                      & \#4 shortest   & \#4 longest    \\
    \hline
    \Nuc{He}{5}{\Lambda} & p & p & 3n  & 144.7 & $>$4.7 &$>$6.0\\ 
                         & p & d & 2n  & 146.9 & $>$8.3 &$>$8.5\\ 
                         & p & d & $\pi^0$ + 2n & 12.0 & $>$8.1 &$>$8.4\\
                         & p & t & $\pi^0$ + n &  18.2 & $>$16.2 &$>$14.1\\
                         & d & p & 2n  & 146.9 & $>$4.7 &$>$6.8\\ 
                         & d & p & $\pi^0$ + 2n & 12.0 & $>$4.7 &$>$6.8\\
                         & d & d & $\pi^0$ + n & 14.2  & $>$9.2 &$>$8.6\\
                         & t & p & $\pi^0$ + n & 18.2  & $>$4.8 &$>$8.6\\
    \hline
    \end{tabular}
\end{table}

\begin{table}[!ht]
    \centering
    \caption{Possible decay modes, their Q-values and kinetic energies at vertex C of E07-T011.\label{tab:T011_C}}
    \begin{tabular}{llllcl}
    \hline
    \#2 & \#6 & \#7 &         & Q-value [MeV]        & $KE_{total}$ [MeV] \\
    \hline
    \Nuc{He}{5}{\Lambda} & p & p & 3n       & 144.7 & $>$75.3 \\
                         & p & d & 2n       & 146.9 & $>$119.0 \\
                         & d & p & 2n       & 146.9 & $>$79.6 \\
                         & d & d & n        & 149.2 & $>$119.9 \\
                         & t & p & n        & 153.2 & $>$91.0 \\
    \hline
    \end{tabular}
\end{table}

\clearpage
\section{Discussion}

\begin{figure}[!ht]
    \centering\includegraphics[width=6.0in]{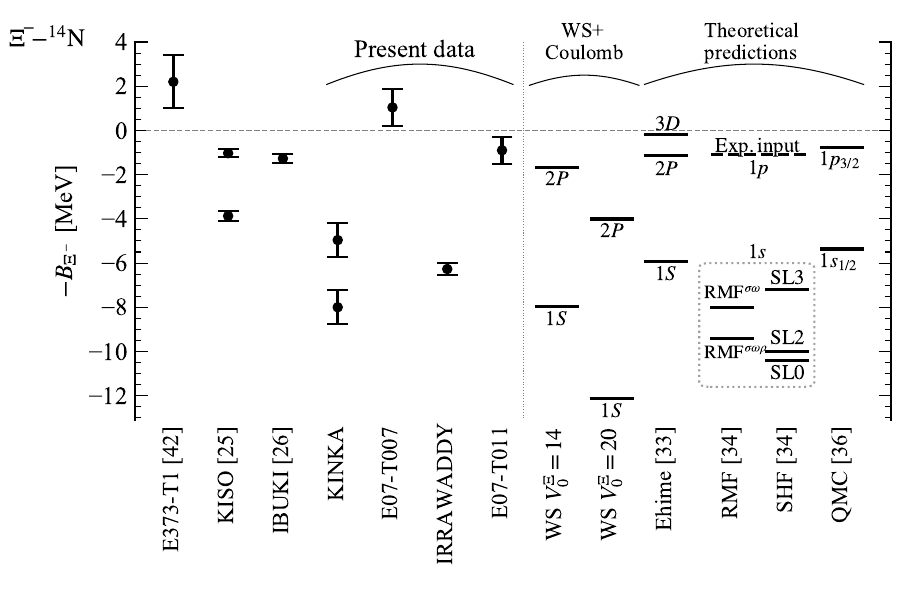}
    \caption{
        $B_{\Xi^-}$ of $\Xi^-$--\Nuc{N}{14}{} system of the E373-T1~\cite{Ichikawa:2001hh}, KISO~\cite{Hiyama:2018lgs}, IBUKI~\cite{Hayakawa:2020oam} events, the present results, 
        theoretical calculations with the WS and Coulomb potential, 
        and theoretical calculations reported in \cite{Yamaguchi:2001ip,Sun:2016tuf,Shyam:2019laf}.
    }\label{fig:BX}
\end{figure}

Table~\ref{tab:N14List} and Fig.~\ref{fig:BX} show the experimental $B_{\Xi^-}$ values of $\Xi^-$--\Nuc{N}{14}{} system for twin hypernuclear events in the E373 and the E07 experiments together with the calculated values.
The theoretically calculated $B_{\Xi^-}$ values taking into account the Coulomb and Ehime (nuclear) potential are 1.14~MeV in the atomic $2P$ (nuclear $1p$) state and 5.93~MeV in the atomic $1S$ (nuclear $1s$) state, respectively~\cite{Yamaguchi:2001ip}.
The $B_{\Xi^-}$ in the $3D$ atomic level is 0.17~MeV, which is determined almost entirely by the Coulomb potential.
The $B_{\Xi^-}$ values based on the RMF and the SHF theories were calculated using $B_{\Xi^-}$=1.1~MeV of the KISO event as the nuclear $1p$ state, ignoring the spin-orbit splitting~\cite{Sun:2016tuf}. 
The $B_{\Xi^-}$ values of the WS are calculated using the Coulomb potential and the WS potential as $\Xi$-nucleus potential where the depth parameter ($V_0^\Xi$) is 14 or 20~MeV.
The $B_{\Xi^-}$ values of the E373-T1, KISO and IBUKI events were reported in Refs.~\cite{Ichikawa:2001hh,Hiyama:2018lgs,Hayakawa:2020oam}. 
Two values of $B_{\Xi^-}$ (8.00 $\pm$ 0.77 and 4.96 $\pm$ 0.77) were derived for the KINKA event, whose daughter \Nuc{Be}{9}{\Lambda} is in the ground state and the excited state, respectively.
The $B_{\Xi^-}$ values of the E07-T007, IRRAWADDY, and E07-T011 were uniquely determined to be $-$1.04 $\pm$ 0.85~MeV, 6.27 $\pm$ 0.27~MeV, and 0.90 $\pm$ 0.62~MeV, respectively.

The KISO and IBUKI events are interpreted in two ways, as discussed in Ref.~\cite{Hayakawa:2020oam}.
In the case (A), we considered that the KISO event has $B_{\Xi^-}$ of 1.03 $\pm$ 0.18~MeV and corresponds to the same $1p$ state as the IBUKI (1.27 $\pm$ 0.21~MeV) event.
Thus, the $1p$ state has the $B_{\Xi^-}$ of 1.13 $\pm$ 0.14~MeV (the weighted average for the two events), and may have a narrow natural width less than 1~MeV.
In the case (B), we considered that the KISO event has $B_{\Xi^-}$ of 3.87 $\pm$ 0.21~MeV and comes from the same $1p$ state as the IBUKI (1.27 $\pm$ 0.21~MeV) event.
Assuming that the two events correspond the same member of the $1p$ spin-orbit doublet, or assuming that the spin-orbit splitting of the $1p$ state is small enough, the $1p$ state has a wide natural width of a few MeV.
In addition, we considered an interpretation (C); the KISO (3.87 $\pm$ 0.21~MeV) and IBUKI (1.27 $\pm$ 0.21~MeV) events are assigned to the two members of the $1p$ spin-orbit doublet, respectively.

In the case (A), the $B_{\Xi^-}$ values of the IRRAWADDY and KINKA events are significantly larger than that of the $1p$ state.
Thus, these events are attributed to the $1s$ state, namely the ground-state spin-doublet ($3/2^+$, $1/2^+$) of the \Nuc{C}{15}{\Xi} hypernucleus.
This is the first observation of the $\Xi$ hypernuclear $1s$ state. 
The weighted averages of $B_{\Xi^-}$ of ``IRRAWADDY and the larger KINKA'' or ``IRRAWADDY and the smaller KINKA'' are obtained to be 6.13 $\pm$ 0.25~MeV or 6.46 $\pm$ 0.25~MeV, respectively.
The two cases of the IRRAWADDY and KINKA pair have a $B_{\Xi^-}$ discrepancy of 1.7 $\pm$ 0.8~MeV or 1.3 $\pm$ 0.8~MeV, respectively.
If the two events correspond to the same member of the $1s$ spin doublet, or if the spin-spin splitting is small enough, the discrepancy indicates that the $1s$ state has a wider natural width than that of the $1p$ state.
If the natural width of the $1s$ state is narrow, the IRRAWADDY and KINKA are assigned to the $1s$ spin doublet.
In the case, the $B_{\Xi^-}$ of the ground state would be around 6.3~MeV or 8.0~MeV.

In the case (B), 
assuming a wide natural width in the $1p$ state,
the $\Xi$ hypernuclear transition from the $1p$ to $1s$ state is strongly suppressed.
Therefore, the IRRAWADDY event and the two cases of the KINKA event would be in the $1p$ state.
Whichever combination is adopted to the $1p$ spin doublet, there must be a $1p$ state with a central $B_{\Xi^-}$ of $\geq$4~MeV.
A $V_0^\Xi$ value of 20~MeV is obtained for a $B_{\Xi^-}$ of 4~MeV in the $1p$ state with the imaginary potential depth ($W_0^\Xi$) setting to be 0~MeV as shown in Fig.~\ref{fig:BX}.
As the $W_0^\Xi$ value increases, the $V_0^\Xi$ value increases to keep $B_{\Xi^-}$ at 4~MeV.
However, the large $V_0^\Xi$ is inconsistent with the results of the BNL E885 experiment~\cite{Khaustov:1999bz}, and consequently the case (B) is unlikely.
Hiyama \textit{et al.}~\cite{Hiyama:2019kpw} calculated the bound system of the light $s$-shell $\Xi$ hypernucleus using the potential of the HAL QCD~\cite{Sasaki:2019qnh}.
The decay width corresponding to the $\Lambda\Lambda$-$\Xi N$ coupling was estimated, and then the decay width in the $S$-wave was obtained to be less than 0.1~MeV with the HAL QCD and less than 1~MeV with Nijmegen ESC08c.
Since the decay width of the $P$-wave is usually equal to, or smaller than, that of the $S$-wave~\cite{hiyama2021}, the case of (B) with a wide natural width in the $1p$ state seems unlikely.

In case (C), it is reasonable to assume that the natural width of each member of the $1p$ spin-orbit doublet is narrow.
The KINKA (4.96 $\pm$ 0.77~MeV) event can be interpreted as a $1p$ state because it is consistent with the $1p$ state of the KISO (3.87 $\pm$ 0.21~MeV) event, within 1.4$\sigma$ of error.
However, the two cases of the KINKA (8.00 $\pm$ 0.77~MeV) and IRRAWADDY (6.27 $\pm$ 0.27~MeV) events are significantly larger than that of the KISO (3.87 $\pm$ 0.21~MeV) event.
Hence, the two cases are most easily interpreted as the $1s$ state of the \Nuc{C}{15}{\Xi} hypernucleus.

Although the $B_{\Xi^-}$ values of the E373-T1 and E07-T007 events are negative, the values are consistent with $B_{\Xi^-}=$0.17~MeV expected for the $3D$ atomic level within 2.0$\sigma$ and 1.4$\sigma$ errors, respectively.
Since the $B_{\Xi^-}$ of the E07-T011 event has a large error, the E07-T011 event can be attributed to the nuclear $1p$ state or the atomic $3D$ state.

The calculated $B_{\Xi^-}$ of 5.93~MeV with the Ehime potential in the nuclear $1s$ state is pretty close to the weighted averages of $B_{\Xi^-}$ being 6.13 $\pm$ 0.25~MeV or 6.46 $\pm$ 0.25~MeV. 
Therefore, this experimental result supports the Ehime potential.
The $B_{\Xi^-}$ value in the nuclear $1s$ state in RMF theory is 9.4~MeV with isospin-dependent potential and 8.0~MeV without that potential, and the latter figure is closer. 
The $B_{\Xi^-}$ with a parameter ${\rm SL_3}$ in SHF theory matches with our results. 
It is to be noted that the present result is almost consistent with the predicted $B_{\Xi^-}$ in a $1s_{1/2}$ state, 5.37~MeV by the QMC model.
The $B_{\Xi^-}$ values by chiral EFT of 5.40~MeV in the $0s$ state and 0.67~MeV in the $0p$ state are almost consistent with the present result. 
Yamamoto has calculated the $B_{\Xi^-}$ values in the ESC16 model~\cite{Nagels:2015lfa} and with the result by HAL QCD~\cite{Sasaki:2019qnh} in Ref.~\cite{Yamamoto:2020ge}.
The $B_{\Xi^-}$ values by ESC16 are 5.7~MeV in the $1s$ state and 1.2~MeV in the $1p$ state and 
the $B_{\Xi^-}$ values by HAL QCD are 5.5~MeV in the $1s$ state and 0.7~MeV in the $1p$ state, respectively.
The values are consistent with the present result, and it should be noted that the latter was obtained from lattice calculations, without use of the experimental data.

Zhu \textit{et al.} calculated a $\Xi^-$ hyperon capture probability by \Nuc{N}{14}{} in the $s$ orbit ($\mathcal{P}_{^{14}N}$) and found it to be 0.00--0.03\%~\cite{Zhu:1991zq}.
Koike reported that a calculated $\mathcal{P}_{^{14}N}$ in the atomic $1S$ (nuclear $1s$) state is 0.0001--0.02\%~\cite{Koike:2017dsh}.
In these calculations, the $W_0^\Xi$ was 0.5--8~MeV.
To compare the calculated $\mathcal{P}_{^{14}N}$ to the experimental results, we estimated the number of $\Xi^-$ absorption in nitrogen detected in the emulsion-counter hybrid method so far. 
In the E176, the E373 and the E07 experiments, 52, 432, and about 2200 events of $\sigma$-stops of $\Xi^-$ hyperon were detected~\cite{Aoki:2009pvs,Theint:2019wkg}.
The percentages of $\Xi^-$ capture in light and heavy nuclei were measured to be 43.2 $\pm$ 3.8\% and 56.8 $\pm$ 4.6\%, respectively~\cite{Theint:2019wkg}.
The ratio of the light nuclei in nuclear emulsion, carbon, nitrogen, and oxygen, was 0.55 : 0.16 : 0.29. 
Even if the $\Xi^-$-absorptive probability by the light nuclei is proportional to the $n$-th power of $Z$ ($0\leq n\leq 5$) and the atomic ratio, 
the probability in nitrogen remains unchanged at 16--17\%.
The $\sigma$-stop formation probability is measured to be about two-thirds of the stopped $\Xi^-$~\cite{Aoki:2009pvs}.
Assuming the formation probability of $\sigma$-stop in $\Xi^-$ capture by each nucleus is the same, about $(52+432+2200)\times 0.43 \times 0.17 \times \frac{3}{2}=300$ stopped $\Xi^-$ must be captured by nitrogen.
If we assume one twin hypernuclear event, IRRAWADDY, is in the $1s$ state, the present result shows that $\mathcal{P}_{^{14}N}$ in the $1s$ state is 0.33\% and the lower limit is 0.035\% at the 90\% confidence level, when a binomial distribution with 300 trials was employed.
The obtained $\mathcal{P}_{^{14}N}$ for the deep orbit is larger than the theoretical estimations by Zhu \textit{et al.} and Koike.
Hence, the $W_0^\Xi$ should be much smaller than 1~MeV, and consequently the $\Xi N$-$\Lambda\Lambda$ coupling is likely weak, which supports the recent lattice QCD calculation~\cite{Sasaki:2019qnh}.

Through the E176, E373, and E07 experiments, the number of twin hypernuclear events uniquely identified to be formed by carbon, nitrogen, and oxygen is 4, 7, and 1, respectively, when two unpublished events consistent with formation in the atomic $3D$ state, one by $\Xi^-$--\Nuc{N}{14}{} and one by $\Xi^-$--\Nuc{O}{16}{}, are included. 
The KISO event detected in the overall scanning method is excluded from this statistics.
The probability of twin hypernuclear formation by $\Xi^-$--\Nuc{N}{14}{} is higher than 16--17\%, which is possibly 
because the decay daughters are based on $\alpha$-clusters; {\it i.e.}, the \Nuc{He}{5}{\Lambda} on $\alpha+\Lambda$, \Nuc{Be}{9}{\Lambda} on 2$\alpha+\Lambda$, and \Nuc{Be}{10}{\Lambda} on 2$\alpha+$n$+\Lambda$ are expected to have large formation probabilities in the fragmentation.

\begin{table}[!ht]
    \centering
    \caption{\label{tab:N14List}Value of $B_{\Xi^-}$ of $\Xi^-$--\Nuc{N}{14}{} for all twin hypernuclei found in the E373 and the E07 experiments. The E373-T1 reflects the updated mass of the $\Xi^-$ hyperon.
    }
    \begin{tabular}{lllS[table-format=2.2,table-figures-uncertainty=1]cS[table-format=2.2,table-figures-uncertainty=1]}
    \hline
    Experiment & Event & Daughters & {$B_{\Xi^-}$ [MeV]} && \\
    \hline
    \hline
    E373 & T1~\cite{Ichikawa:2001hh}            & \Nuc{He}{5}{\Lambda}+\Nuc{He}{5}{\Lambda}+\Nuc{He}{4}{}+n & -2.2+-1.2&&               \\
    E373 & T2 KISO~\cite{Hiyama:2018lgs}        & \Nuc{Be}{10}{\Lambda} + \Nuc{He}{5}{\Lambda}    & 3.87+-0.21 &or& 1.03+-0.18 \\
    E07  & T006 IBUKI~\cite{Hayakawa:2020oam}    & \Nuc{Be}{10}{\Lambda} + \Nuc{He}{5}{\Lambda}    & 1.27+-0.21&&              \\
    \hline
    E373 & T3 KINKA                             & \Nuc{Be}{9}{\Lambda}+\Nuc{He}{5}{\Lambda}+n     & 8.00+-0.77 &or& 4.96+-0.77\\
    E07  & T007                                  & \Nuc{Be}{9}{\Lambda}+\Nuc{He}{5}{\Lambda}+n     & -1.04+- 0.85&&              \\
    E07  & T010 IRRAWADDY                        & \Nuc{He}{5}{\Lambda}+\Nuc{He}{5}{\Lambda}+\Nuc{He}{4}{}+n & 6.27+-0.27&&              \\
    E07  & T011                                      & \Nuc{He}{5}{\Lambda}+\Nuc{He}{5}{\Lambda}+\Nuc{He}{4}{}+n & 0.90+-0.62               \\
    \hline
    \end{tabular}
\end{table}

\clearpage
\section{Summary and prospects}
The study of the $\Xi N$ interaction is advancing with a increasing number of $\Xi$ hypernuclear events and accurate binding energies measurements of the $\Xi^-$ hyperon and a nucleus ($B_{\Xi^-}$) found in the nuclear emulsion.
Four clear twin hypernuclear events, E373-T3 (KINKA), E07-T007, E07-T010 (IRRAWADDY) and E07-T011 were found.
The KINKA event was identified to be $\Xi^-$ + \Nuc{N}{14}{} $\to$ \Nuc{Be}{9}{\Lambda} + \Nuc{He}{5}{\Lambda} + n and the $B_{\Xi^-}$ value was found to be 8.00 $\pm$ 0.77 or 4.96 $\pm$ 0.77~MeV.
The IRRAWADDY event was identified to be $\Xi^-$ + \Nuc{N}{14}{} $\to$ \Nuc{He}{5}{\Lambda} + \Nuc{He}{5}{\Lambda} + \Nuc{He}{4}{} + n and the $B_{\Xi^-}$ value was found to be 6.27 $\pm$ 0.27~MeV.
The $B_{\Xi^-}$'s of the IRRAWADDY and KINKA events are significantly deeper than the nuclear $1p$ state with a narrow natural width, suggested by the KISO (1.03 $\pm$ 0.18~MeV) and IBUKI (1.27 $\pm$ 0.21~MeV) events. 
The two events seem to show the nuclear $1s$ state of \Nuc{C}{15}{\Xi} with a weighted average of $B_{\Xi^-} $ being 6.13 $\pm$ 0.25~MeV or 6.46 $\pm$ 0.25~MeV.
Even if the KISO (3.87 $\pm$ 0.21~MeV) event is assigned to the $1p$ state, assuming a narrow natural width, 
the $B_{\Xi^-}$ of the IRRAWADDY (6.27 $\pm$ 0.27~MeV) and KINKA (8.00 $\pm$ 0.77~MeV) events are deeper than the $1p$ state.
Assuming that the $1p$ state has a wide natural width from KISO (3.87 $\pm$ 0.21~MeV) to IBUKI, 
the IRRAWADDY and KINKA events are attributed to the $1p$ state due to the low transition probability of $\Xi^-$ hyperon from the $1p$ to the $1s$ state.
The deep $1p$ state around 4~MeV is in contradiction with the results of the BNL E885 experiment.
We concluded, therefore, the $B_{\Xi^-}$ of the IRRAWADDY and KINKA events leads to the first observation of a nuclear $1s$ state of the $\Xi$ hypernucleus, \Nuc{C}{15}{\Xi}.

To clarify the energy and the width of the $1s$ and the $1p$ states, it is expected to find more events like the IRRAWADDY event with a unique interpretation of the decay modes and the small error in $B_{\Xi^-}$.
Such events would provide more precise information about the $\Xi^-$--\Nuc{N}{14}{} system.

Although the ``emulsion-counter hybrid method'' has been successful in detecting 2 twin hypernuclei in the E373 and 13 twin hypernuclei in the E07 experiments so far,
numerous events remain to be identified in the emulsion sheets.
To search for these unidentified events, an overall scanning method has been initiated and the analysis over the entire volume of the nuclear emulsion has begun. 
Through this analysis, 
10 times more twin hypernuclear events will be acquired, and then richer information will be obtained regarding the $\Xi N$ interaction.

\clearpage
\section{Acknowledgments}

We thank the KEK staff for the E373 experiment.
We also thank the staff of the J-PARC accelerator and the Hadron experimental facility, FUJIFILM Corporation, KEK computing research, NII for SINET5, Kamioka Observatory, ICRR for the E07 experiment.
We thank E.~Hiyama for the useful discussion on the interpretation.
This work was supported by 
Japan Society for the Promotion of Science (JSPS) KAKENHI Grant Numbers 
JP23224006, 
JP16H02180, 
JP20H00155, 
JP20J00682, 
Ministry of Education, Culture, Sports, Science and Technology (MEXT) KAKENHI Grant Numbers 
JP15001001, 
JP24105002, 
JP18H05403, 
JP19H05147, 
National Research Foundation (NRF) of Korea with Grant Number 2018R1A2B2007757.
J.P. 
is supported by Deutscher Akademischer Austauschdienst (DAAD) 
Grant numbers PPP Japan 2017, No. 57345296 by DE/Federal Republic of Germany and
Grant agreement STRONG 2020, No. 824093 by EU/European Union.

\appendix
\def\thesection{Appendix \Alph{section}}
\section{\label{app:mass}Mass and excitation energy}

Table \ref{tab:SLH} shows the rest mass of single $\Lambda$ hypernuclei and $\Xi^-$ hyperon and the excitation energy used in this paper.
The $B_\Lambda$ values of \Nuc{He}{5}{\Lambda} and \Nuc{Be}{9}{\Lambda} were measured to be 3.12 $\pm$ 0.02~MeV and 6.71 $\pm$ 0.04~MeV, respectively using the nuclear emulsion~\cite{Juric:1973zq}.
The $B_\Lambda$ of \Nuc{Be}{10}{\Lambda} was measured to be 8.55 $\pm$ 0.13~MeV using the missing mass spectroscopy~\cite{Gogami:2015tvu}.
The excitation energy of \Nuc{Be}{9}{\Lambda} is an average of $\gamma$-ray energies of two E2 transitions (5/2$^+$$\to$1/2$^+$ and 3/2$^+$$\to$1/2$^+$)~\cite{Akikawa:2002tm}.
The excitation energy of \Nuc{Be}{10}{\Lambda} is the first separable gap of $B_\Lambda$ values, because the first excited state and the ground state are a doublet with energy spacing of $\sim$0.1~MeV~\cite{Gogami:2015tvu}.

\begin{table}[!ht]
    \centering
    \caption{\label{tab:SLH}Rest mass of single $\Lambda$ hypernuclei and $\Xi^-$ hyperon and the excitation energy.}
    \begin{tabular}{lS[table-format=4.2,table-figures-uncertainty=1]}
    \hline
                         & {Rest mass in ground state [MeV]}   \\
    \hline
    \Nuc{He}{5}{\Lambda}~\cite{Juric:1973zq}    & 4839.94+-0.02   \\ 
    \Nuc{Be}{9}{\Lambda}~\cite{Juric:1973zq}    & 8563.82+-0.04   \\ 
    \Nuc{Be}{10}{\Lambda}~\cite{Gogami:2015tvu} & 9499.88+-0.13   \\ 
    $\Xi^-$   & 1321.71+-0.07   \\
    \hline
                                & {Excitation energy [MeV]}    \\
    \hline
    \Nuc{Be}{9}{\Lambda}~\cite{Akikawa:2002tm}  & 3.04+-0.02 \\
    \Nuc{Be}{10}{\Lambda}~\cite{Gogami:2015tvu} & 2.78+-0.11   \\
    \hline
    \end{tabular}
\end{table}

\clearpage
\bibliographystyle{ptephy}
\bibliography{xihpaper}

\begin{thebibliography}{10}

\bibitem{Jaffe:1976yi}
R.L. Jaffe, Phys. Rev. Lett., {\bf 38}, 195--198, [Erratum: Phys.Rev.Lett. 38,
  617 (1977)] (1977).

\bibitem{SchaffnerBielich:2008kb}
Jürgen Schaffner-Bielich, Nucl. Phys. A, {\bf 804}, 309--321 (2008),
  {{arXiv:0801.3791}}.

\bibitem{Davis:2005mb}
D.H. Davis, Nucl. Phys. A, {\bf 754}, 3--13 (2005).

\bibitem{Hashimoto:2006aw}
O.~Hashimoto and H.~Tamura, Prog. Part. Nucl. Phys., {\bf 57}, 564--653 (2006).

\bibitem{Gal:2016boi}
A.~Gal, E.~V. Hungerford, and D.~J. Millener, Rev. Mod. Phys., {\bf 88}(3),
  035004 (2016),  {{arXiv:1605.00557}}.

\bibitem{Danysz:1963zza}
M.~Danysz et~al., Nucl. Phys., {\bf 49}, 121--132 (1963).

\bibitem{Mondal:1979hp}
A.S. Mondal, A.K. Basak, M.M. Kasim, and A.~Husain, Nuovo Cim. A, {\bf 54},
  333--339 (1979).

\bibitem{Dover:1982ng}
C.B. Dover and A.~Gal, Annals Phys., {\bf 146}, 309--348 (1983).

\bibitem{Aoki:1998sv}
S.~Aoki et~al., Nucl. Phys. A, {\bf 644}, 365--385 (1998).

\bibitem{Aoki:1991ip}
S.~Aoki et~al., Prog. Theor. Phys., {\bf 85}, 1287--1298 (1991).

\bibitem{Aoki:2009pvs}
S.~Aoki et~al., Nucl. Phys. A, {\bf 828}, 191--232 (2009).

\bibitem{10.1143/ptp/89.2.493}
S.~Aoki et~al., Prog. Theor. Phys., {\bf 89}(2), 493--500 (1993).

\bibitem{Aoki:1995za}
S.~Aoki et~al., Phys. Lett. B, {\bf 355}, 45--51 (1995).

\bibitem{yamamoto1994formation}
Y.~Yamamoto, T.~Motoba, T.~Fukuda, M.~Takahashi, and K.~Ikeda, Prog. Theor.
  Phys. Suppl., {\bf 117}, 281--306 (1994).

\bibitem{Fukuda:1998bi}
T.~Fukuda et~al., Phys. Rev. C, {\bf 58}, 1306--1309 (1998).

\bibitem{Khaustov:1999bz}
P.~Khaustov et~al., Phys. Rev. C, {\bf 61}, 054603 (2000),
  {{nucl-ex/9912007}}.

\bibitem{Ichikawa:1998yk}
A.~Ichikawa et~al., Nucl. Instrum. Meth. A, {\bf 417}, 220--229 (1998).

\bibitem{ProposalE07}
K.~Imai, K.~Nakazawa, and H.~Tamura, Proposals for Nuclear and Particle Physics
  Experiments at J-PARC, {\bf 1} (2006).

\bibitem{Theint:2019wkg}
A.M.M. Theint et~al., Prog. Theor. Exp. Phys., {\bf 2019}(2), 021D01 (2019).

\bibitem{Takahashi:2001nm}
H.~Takahashi et~al., Phys. Rev. Lett., {\bf 87}, 212502 (2001).

\bibitem{Ahn:2013poa}
J.K. Ahn et~al., Phys. Rev. C, {\bf 88}(1), 014003 (2013).

\bibitem{Ekawa:2018oqt}
H.~Ekawa et~al., Prog. Theor. Exp. Phys., {\bf 2019}(2), 021D02 (2019),
  {{arXiv:1811.07726}}.

\bibitem{Yoshida:2017oww}
J.~Yoshida et~al., Nucl. Instrum. Meth. A, {\bf 847}, 86--92 (2017).

\bibitem{Nakazawa:2015joa}
K.~Nakazawa et~al., Prog. Theor. Exp. Phys., {\bf 2015}(3), 033D02 (2015).

\bibitem{Hiyama:2018lgs}
E.~Hiyama and K.~Nakazawa, Ann. Rev. Nucl. Part. Sci., {\bf 68}, 131--159
  (2018).

\bibitem{Hayakawa:2020oam}
S.~H. Hayakawa et~al., Phys. Rev. Lett., {\bf 126}(6), 062501 (2021),
  {{arXiv:2010.14317}}.

\bibitem{Batty:1998fn}
C.~J. Batty, E.~Friedman, and A.~Gal, Phys. Rev. C, {\bf 59}, 295--304 (1999),
  {{nucl-th/9809042}}.

\bibitem{Nagels:1976xq}
M.~M. Nagels, T.~A. Rijken, and J.~J. de~Swart, Phys. Rev. D, {\bf 15}, 2547
  (1977).

\bibitem{Nagels:1978sc}
M.~M. Nagels, T.~A. Rijken, and J.~J. de~Swart, Phys. Rev. D, {\bf 20}, 1633
  (1979).

\bibitem{Maessen:1989sx}
P.~M.~M. Maessen, T.~A. Rijken, and J.~J. de~Swart, Phys. Rev. C, {\bf 40},
  2226--2245 (1989).

\bibitem{Rijken:1998yy}
T.~A. Rijken, V.~G.~J. Stoks, and Y.~Yamamoto, Phys. Rev. C, {\bf 59}, 21--40
  (1999),  {{nucl-th/9807082}}.

\bibitem{Nagels:2015lfa}
M.~M. Nagels, Th.~A. Rijken, and Y.~Yamamoto, Phys. Rev. C, {\bf 99}(4), 044003
  (2019),  {{arXiv:1501.06636}}.

\bibitem{Yamaguchi:2001ip}
M.~Yamaguchi, K.~Tominaga, T.~Ueda, and Y.~Yamamoto, Prog. Theor. Phys., {\bf
  105}, 627--648 (2001).

\bibitem{Sun:2016tuf}
T.T. Sun, E.~Hiyama, H.~Sagawa, H.J. Schulze, and J.~Meng, Phys. Rev. C, {\bf
  94}(6), 064319 (2016),  {{arXiv:1611.03661}}.

\bibitem{Guichon:2008zz}
P.A.M. Guichon, A.W. Thomas, and K.~Tsushima, Nucl. Phys. A, {\bf 814}, 66--73
  (2008),  {{arXiv:0712.1925}}.

\bibitem{Shyam:2019laf}
R.~Shyam and K.~Tsushima (2019),  {{arXiv:1901.06090}}.

\bibitem{Polinder:2007mp}
H.~Polinder, J.~Haidenbauer, and U.~G. Meissner, Phys. Lett. B, {\bf 653},
  29--37 (2007),  {{arXiv:0705.3753}}.

\bibitem{Kohno:2019oyw}
M.~Kohno, Phys. Rev. C, {\bf 100}(2), 024313 (2019),  {{arXiv:1908.01934}}.

\bibitem{Ishii:2006ec}
N.~Ishii, S.~Aoki, and T.~Hatsuda, Phys. Rev. Lett., {\bf 99}, 022001 (2007),
  {{nucl-th/0611096}}.

\bibitem{Sasaki:2019qnh}
K.~Sasaki et~al., Nucl. Phys. A, {\bf 998}, 121737 (2020),
  {{arXiv:1912.08630}}.

\bibitem{Fujita:2020}
Manami Fujita,
\newblock {\em {Experimental study for spectroscopy of $\Xi$--atomic X rays}},
\newblock PhD thesis, Tohoku U. (2020).

\bibitem{Ichikawa:2001hh}
A.~Ichikawa et~al., Phys. Lett. B, {\bf 500}, 37--46 (2001).

\bibitem{Hayakawa:2019dth}
Shuhei Hayakawa,
\newblock {\em {Study of~ $\Xi$-nucleus interaction by measurement of twin
  hypernuclei with hybrid emulsion method}},
\newblock PhD thesis, Osaka U. (2019).

\bibitem{Ekawa:2020fya}
Hiroyuki Ekawa,
\newblock {\em {Observation of a Double-\ensuremath{\Lambda} Hypernucleus
  \ensuremath{\Lambda}\ensuremath{\Lambda}Be with Hybrid Emulsion Method at
  J-PARC}},
\newblock PhD thesis, Kyoto U. (2020).

\bibitem{MyintKyawSoe:2017gpq}
Myint~Kyaw Soe et~al., Nucl. Instrum. Meth. A, {\bf 848}, 66--72 (2017).

\bibitem{Juric:1973zq}
M.~Jurič et~al., Nucl. Phys. B, {\bf 52}, 1--30 (1973).

\bibitem{Bertini:1981zx}
R.~Bertini et~al., Nucl. Phys. A, {\bf 368}, 365--374 (1981).

\bibitem{May:1981er}
M.~May et~al., Phys. Rev. Lett., {\bf 47}, 1106--1109 (1981).

\bibitem{Dluzewski:1988ye}
P.~Dłużewski, K.~Garbowska-Pniewska, J.~Pniewski, T.~Tymieniecka, P.~Ciok,
  and D.~H. Davis, Nucl. Phys. A, {\bf 484}, 520--524 (1988).

\bibitem{Davis:1986kg}
D.~H. Davis and J.~Pniewski, Contemp. Phys., {\bf 27}, 91--116 (1986).

\bibitem{Cusanno:2008xx}
F.~Cusanno et~al., Phys. Rev. Lett., {\bf 103}, 202501 (2009),
  {{arXiv:0810.3853}}.

\bibitem{Botta:2012xi}
E.~Botta, T.~Bressani, and G.~Garbarino, Eur. Phys. J. A, {\bf 48}, 41 (2012),
  {{arXiv:1203.5707}}.

\bibitem{Gogami:2015tvu}
T.~Gogami et~al., Phys. Rev. C, {\bf 93}(3), 034314 (2016),
  {{arXiv:1511.04801}}.

\bibitem{Avery:1991}
P.~Avery,
\newblock {\em Applied Fitting Theory I: General Least Squares Theory, CLEO
  Note CBX 91-72},
\newblock  (\url{https://www.phys.ufl.edu/~avery/fitting.html}, 1991).

\bibitem{Kinbara:2019kyx}
S.~Kinbara et~al., Prog. Theor. Exp. Phys., {\bf 2019}(1), 011H01 (2019).

\bibitem{Akikawa:2002tm}
H.~Akikawa et~al., Phys. Rev. Lett., {\bf 88}, 082501 (2002).

\bibitem{Hiyama:2019kpw}
E.~Hiyama, K.~Sasaki, T.~Miyamoto, T.~Doi, T.~Hatsuda, Y.~Yamamoto, and Th.~A.
  Rijken, Phys. Rev. Lett., {\bf 124}(9), 092501 (2020),  {{arXiv:1910.02864}}.

\bibitem{hiyama2021}
Emiko Hiyama,
\newblock private communication (2021).

\bibitem{Yamamoto:2020ge}
Y.~Yamamoto, Genshikaku Kenkyu, {\bf 65-1}, 54 (2020).

\bibitem{Zhu:1991zq}
D.~Zhu, C.~B. Dover, A.~Gal, and M.~May, Phys. Rev. Lett., {\bf 67}, 2268--2271
  (1991).

\bibitem{Koike:2017dsh}
Takahisa Koike, JPS Conf. Proc., {\bf 17}, 033011 (2017).

\end{thebibliography}

\end{document}